\documentclass[%
 reprint,
nofootinbib,
longbibliography, 
 amsmath,amssymb,
prstab,
floatfix,
]{revtex4-1}

\usepackage{graphicx}% Include figure files
\usepackage{dcolumn}% Align table columns on decimal point
\usepackage{bm}% bold math
\usepackage{color}
\newcommand{\enum}[2]{\ensuremath{#1\times10^{#2}}} % a number in scientific notation
\newcommand{\qty}[2]{\ensuremath{#1\,\mathrm{#2}}}  %% quantity as number and unit

\newcommand{\elumi}[2]{\mbox{\qty{\enum{#1}{#2}}{cm^{-2}s^{-1}}}}

\newcommand{\Nb}{\ensuremath{N_b}}

\newcommand{\lumival}[2]{$ L = \elumi{#1}{#2}$}

\newcommand{\isotope}[3]{\ensuremath{^{#1}\mathrm{#2}^{#3}}}
\newcommand{\Pb}{\isotope{208}{Pb}{82+}}
\newcommand{\PbBFPP}{\isotope{208}{Pb}{81+}}
\newcommand{\Eb}{\ensuremath{E_b}}
\newcommand{\sigmaBFPP}{\ensuremath{\sigma_{\text{BFPP}}}}
\newcommand{\PBFPP}{\ensuremath{P_{\text{BFPP}}}}

\newcommand{\PbPb}{Pb-Pb}  % should definitely not be an en-dash (--).  Maybe no dash at all. 
\newcommand{\pPb}{p-Pb} 
\newcommand{\pp}{p-p} 
\newcommand{\lRF}{\lambda_{\text{RF}}}

\begin{document}

\preprint{APS/123-QED}

%%JMJ%% I propose the following alternative title which puts more emphasis on the quench limit and could increase the readership and impact of the paper.   What do you think ?
%%
\title{Bound-free pair production from nuclear collisions\\ and the steady-state quench limit
of the main dipole magnets \\of 
the CERN Large Hadron Collider     
%at 6.37~Z~TeV per beam
}% Force line breaks with \\
% \title{Bound-free pair production in 
% P\MakeLowercase{b}-P\MakeLowercase{b}  
% operation of the CERN\\ 
% Large Hadron Collider     
%at 6.37~Z~TeV per beam

\author{M.~Schaumann} 
	\email{Michaela.Schaumann@cern.ch} 
\author{J.M.~Jowett} 
\author{C.~Bahamonde~Castro} 
\author{R.~Bruce}
\author{A.~Lechner}
\author{T.~Mertens}
\affiliation{%
 CERN, Geneva, Switzerland
}%

\date{\today}% It is always \today, today,

\begin{abstract} 

During its Run~2 (2015--2018),  the Large Hadron Collider (LHC) operated at almost twice higher energy,  
and provided \PbPb\ 
collisions  with an order of magnitude higher luminosity, 
than in the previous Run~1.   
In consequence, the power of the secondary beams emitted from the interaction points by the
bound-free pair production (BFPP) process increased by a factor $\sim20$, while the propensity of the bending magnets to quench increased with the higher magnetic field.    
This beam power is about 35~times greater than that contained in the luminosity debris from hadronic interactions and is focused on specific locations that fall naturally inside  superconducting magnets.  
The  risk of quenching these magnets has long been recognized as severe and there are   
operational limitations due to the dynamic heat load 
that must be evacuated by the cryogenic system.

High-luminosity operation was nevertheless possible thanks to orbit bumps that were introduced in the dispersion suppressors  around the ATLAS and CMS experiments to prevent quenches by displacing and spreading out these beam losses.  
Further, in 2015, the BFPP beams were manipulated to induce a controlled quench, 
thus providing the first direct measurement of the steady-state quench level of an LHC dipole magnet.  
The same experiment demonstrated the need for new collimators that are being installed 
around the ALICE experiment 
to intercept the secondary beams in the future.  
This paper discusses the experience with BFPP at luminosities very close to the 
future High Luminosity LHC (HL-LHC) target,   gives results on the 
risk reduction by orbit bumps and presents a detailed analysis of the controlled quench experiment.

\end{abstract}

\maketitle

%\tableofcontents

\section{\label{sec:Intro}Introduction}

In its second major physics program, the Large Hadron Collider (LHC)~\cite{lhcdesignV1} operates with nuclear beams to study strongly-interacting matter---notably  the Quark-Gluon Plasma---at the highest temperatures and densities available. 
For about one month at the end of each operational year, the LHC collides fully stripped lead (\Pb) ions with each other or with protons. 
So far four full \PbPb\ runs have been executed in the years 2010, 2011, 2015 and 2018~\cite{ipac11_jowett_fist_Pb_run, ipac2012:Bartosik:2012zza, MSchaumannThesis, ipac2016:PbPb2015, jowett19_evian, Jowett:IPAC2019-WEYYPLM2}
\footnote{Additionally three \pPb\ runs took place in 2012, 2013 and 2016~\cite{ipac2013:Jowett:1572994, ipac2017:Jowett:2289686}.}.
The LHC has four interaction points (IPs) that host the main experiments ATLAS (IP1), ALICE (IP2), CMS (IP5) and LHCb (IP8). Since 2015 all of them have been participating in \PbPb\ data taking\footnote{LHCb was the last experiment to join the heavy-ion community, taking its first ion collisions in the  pilot \pPb\ run in 2012.}. 
Details of the operational conditions and differences between the interaction regions (IRs), as well as achieved luminosities will be given in Section~\ref{sec:Run2}.
Table~\ref{t:beamParOverview} summarises the Pb beam parameters from  the original LHC design, the maximum achieved in operation and those expected for high-luminosity operation in Run~3 (starting in 2022).

%%%%%%%%%%%%%%%%%%%%%%%%%%%%%%%
\begin{table*}
  \centering
    \begin{tabular}{lccc}     
    \hline
    \hline
                & \textbf{~LHC design~}   & \textbf{~75\,ns (2018)~} & \textbf{~HL-LHC} \\ 
    \hline
    
Beam energy [Z\,TeV]    & 7  & 6.37 & 7 \\               
Total number of bunches    & 592          & 733  & 1240 \\ 
Bunch intensity [$10^7$ Pb ions] & 7 & 21 & 18 \\ 
Normalized transverse emittance [$\mu$m] & 1.5 & 2.3 & 1.65           \\
RMS bunch length [cm] &  7.94 & 8.24 &  7.94          \\ 
$\beta^*$ in IP (1/2/5/8) [m] & (0.55 / 0.5 / 0.55 / 10.0)  & (0.5 / 0.5 / 0.5 / 1.5) & (0.5 / 0.5 / 0.5 / 1.5)  \\
Net crossing-angle IP (1/2/5/8) [$\mu$rad] & (160 / 40 / 160 / -) & (160 / 60 / 160 / 320) & (170 / 100 / 170 / 305)  \\
Peak luminosity IP (1/2/5/8) [\qty{}{cm^{-2}s^{-1}}] & (1.0 / 1.0 / 1.0 / -)  & (6.1 / - / 6.1 / -) & -          \\
Levelled luminosity IP (1/2/5/8) [\qty{}{cm^{-2}s^{-1}}] & -  & (- / 1.0 / - / 1.0) & (7.0 / 7.0 / 7.0 / 1.0)          \\
\hline
\hline
    \end{tabular}%
   \caption{Pb beam and main optics parameters at collision in the LHC design report~\cite{lhcdesignV1}, as achieved in 2018~\cite{jowett19_evian,Jowett:IPAC2019-WEYYPLM2}, and as envisaged for HL-LHC~\cite{Bruce:HL-LHC-Update2020}. The 2018 parameters refer to the average typical in the fills with 75~ns bunch spacing.}
  \label{t:beamParOverview}%
\end{table*}
%%%%%%%%%%%%%%%%%%%%%%%%%%%%%%%

\begin{figure}[tbp]
\centering\includegraphics[width= 1\columnwidth]{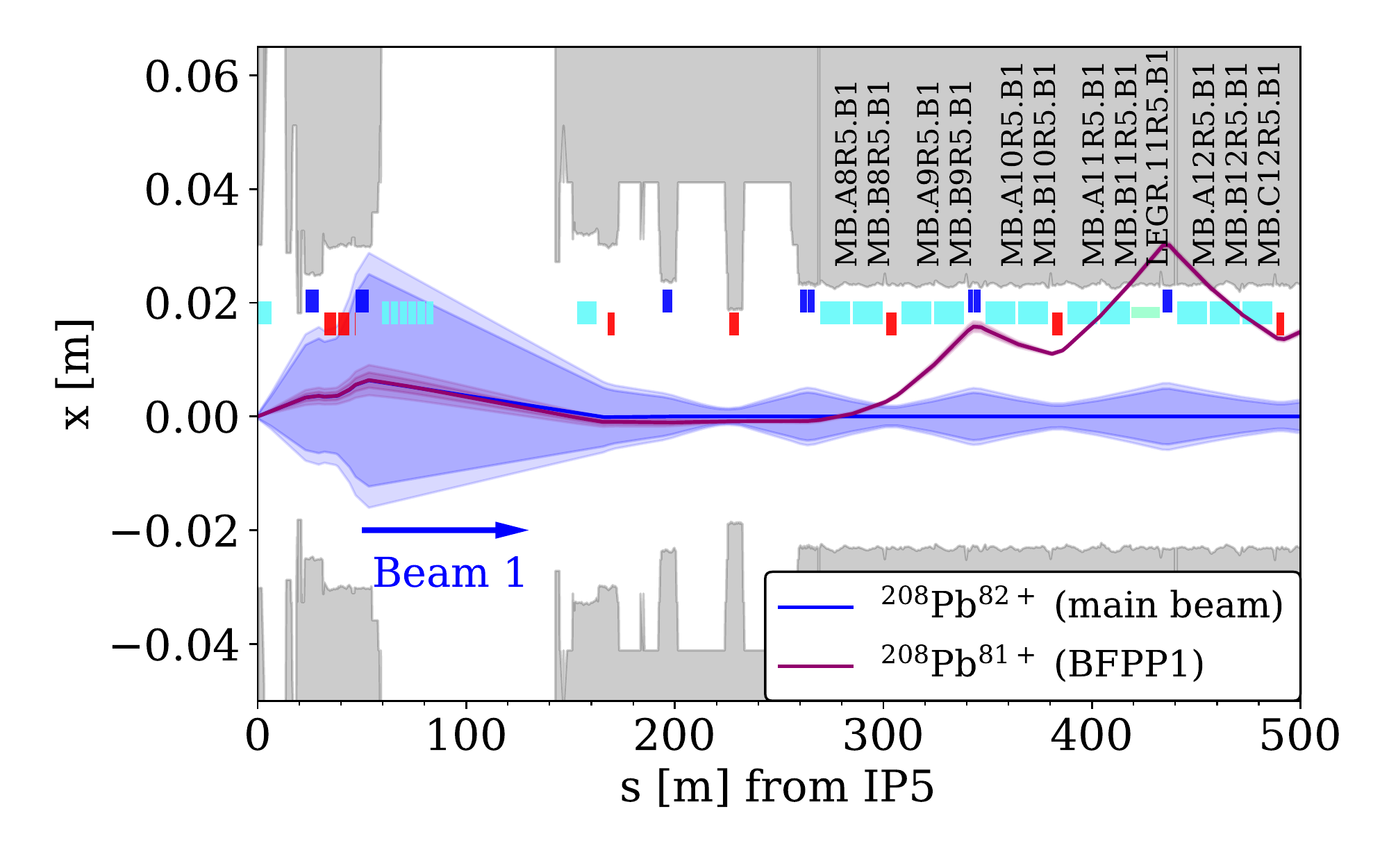}
\vspace{-0.7cm}
\caption{Example of main \Pb~(blue, 10--12$\sigma$) and BFPP1 \PbBFPP~beam (red, 1--2$\sigma$)  envelopes, and aperture (grey) in the horizontal plane, Beam~1 direction right of IP5 (at $s=0$). 
Beam-line elements are indicated schematically as rectangles. Dipoles in light blue, quadrupoles in dark blue (focusing) and red (defocusing). While the main beam travels through the center of the beam-line elements in the dispersion suppressor (starting at about \qty{250}{m}), the BFPP1 beam separates and impacts in the aperture of the second superconducting dipole magnet of cell~11.
}
\label{f:bfppPath}
\end{figure}

Major operational challenges and luminosity limits in \PbPb\ operation of the LHC originate from 
those interactions between lead nuclei in the colliding bunches which have impact parameter $b>2R$, where $R$ is the nuclear radius. 
Since the nuclei do not overlap these \emph{ultra-peripheral}  interactions are purely electromagnetic.  
Among many possible reactions,  two effects dominate: 
(1) copious lepton-pair production in collisions between quasi-real photons, and 
(2) emission of nucleons in electromagnetic dissociation (EMD) of the nuclei, dominated by excitation of the Giant Dipole Resonance. 
Most of the pair production is innocuous except for the (single) bound-free pair production (BFPP1):
\begin{equation*}
\Pb + \Pb \longrightarrow \Pb + \PbBFPP +  e^+,
\end{equation*}
in which the electron is created in a bound state of one nucleus.
Among the EMD processes, the channels where one nucleus loses either one or two neutrons are the most frequent:
\begin{eqnarray*}
\Pb + \Pb \longrightarrow \Pb + ^{207}\mathrm{Pb}^{82+} +  \mathrm{n},
\\
\Pb + \Pb \longrightarrow \Pb + ^{206}\mathrm{Pb}^{82+} +  2\mathrm{n}.
\end{eqnarray*}
As extensively discussed previously (see, e.g., \cite{Klein:2000ba,Jowett:619634,epac2004:Jowett:2004rd,PhysRevLett.99.144801,PhysRevSTAB.12.071002,MSchaumannThesis} and further references therein), the modified nuclei emerge from the interaction point (IP), 
as a narrow secondary beam with modified magnetic rigidity, following a dispersive trajectory. 
This is illustrated in Fig.~\ref{f:bfppPath}, where the red line indicates the trajectory of the BFPP1 ions that separate from the main beam (blue) when they enter the dispersion suppressor (DS) downstream from the IP. 
This beam impacts over just a few meters longitudinally, on the beam screen in a superconducting magnet of the DS, giving rise to a localized power deposition in the magnet coils.
These secondary beams emerge in both directions from every IP where ions collide. 
Each carries a power of  
 \begin{equation}
% \PBFPP = L \cdot \sigmaBFPP \cdot \Eb,
%%  P_p = L \cdot \sigma_p \cdot \Eb,
 P_p = L  \sigma_p  \Eb,
 \label{eq:PBFPP}
 \end{equation}
 where $L$ is the instantaneous luminosity, $\sigma_p$ the interaction cross-section of the corresponding  process  (BFPP1, EMD1, EMD2, etc.) and $E_b$ is the beam energy. At the 2015/18 beam energy of $\Eb=\qty{6.37\,Z}{TeV}$~\cite{ipac2016:PbPb2015, Jowett:IPAC2019-WEYYPLM2}, the 
 theoretical\footnote{
        Measurements are not available at the time of writing.
        } 
 cross-section for the BFPP1 process is   $\sigma_\mathrm{BFPP1} \simeq \qty{276}{b}$~\cite{Meier:2000ga}. 
 Cross-sections for the EMD processes are 
\mbox{$\sigma_\mathrm{EMD1} \simeq\qty{95}{b}$} and
 \mbox{$\sigma_\mathrm{EMD2} \simeq\qty{30}{b}$}~\cite{fluka_BOHLEN2014211,BATTISTONI201510}.
As Eq.~\eqref{eq:PBFPP} shows, these losses carry much greater power than the luminosity debris from 
nuclear interactions of total cross-section \qty{8}{b}.

%%%%%%%%%%%%%%%%%%%%%%%%%%%%%%%
%%%%%%%%%%%%%%%%%%%%%%%%%%%%%%%

Since these effects are directly proportional to the luminosity, they will be of even greater concern after the high-luminosity upgrade that is being implemented in the current Long Shutdown (LS2, 2019-2021). The key upgrades that will have influence on the secondary beam power are the lifting of the  limit on peak luminosity in the ALICE experiment from the current \lumival{1}{27} to about \lumival{7}{27}~\cite{Abelev:1475243}, and the RF upgrade in the Super Proton Synchrotron (SPS) that will reduce the bunch spacing to \qty{50}{ns}~\cite{Argyropoulos:IPAC2019-WEPTS039, Quartullo:2658075} allowing for a higher circulating beam intensity and increased potential peak luminosity in all experiments. 

From a comparison of the interaction cross-sections, it is clear that the BFPP1 secondary beam poses the greatest risk. It carries enough power to quench magnets and directly limit luminosity (see e.g.~\cite{Jowett:IPAC2016-TUPMW028, BahamondeCastro:IPAC2016-TUPMW006}). The quench experiment that will be discussed in detail in Section~\ref{sec:quenchTest} showed that the BFPP beams can quench a superconducting dipole at a luminosity of $L\approx \enum{2.3}{27}\mathrm{cm^{-2} s^{-1}}$, if the full secondary beam impacts directly in the magnet. 
The EMD beams are of less concern, because their power is about 
2.9~times lower than that of the BFPP1 beam. Moreover, the rigidity change of the EMD1 is small enough such that those particles do not impact in the DS, but continue travelling on their dispersive trajectory until they are intercepted by the momentum cleaning collimators.  For those reasons, the  discussions in this paper will concentrate on the BFPP beams and their consequences.

The phenomena discussed in this paper are only significant for the \PbPb\ colliding beam mode. 
The production of secondary beams in \pPb\ collisions is negligible, because of the much reduced interaction cross-section. The BFPP cross-section in \pPb\ collisions is only around \qty{40}{mb}~\cite{marc-thesis} although the corresponding luminosity is two orders of magnitude greater.  
In high luminosity proton-proton collisions, the cross-section, at a few pb, is much smaller still and results in an occasional (\qty{\sim 0.1}{Hz}) multi-TeV neutral hydrogen atom travelling down the center of the beam pipe.

%%%%%%%%%%%%%%%%%%%%%%%%%%%%%%%%%%%%%%%%%%%%%%%%%%%%%%%%%%%%%%%%%%%%%%%%%%%%%%%%%%%%%%%%%%%%%%%%%%
%%%%%%%%%%%%%%%%%%%%%%%%%%%%%%%%%%%%%%%%%%%%%%%%%%%%%%%%%%%%%%%%%%%%%%%%%%%%%%%%%%%%%%%%%%%%%%%%%%

%%%%%%%%%%%%%%%%%%%%%%%%%%%%%%%
\begin{figure*}
\centering\includegraphics[width= \textwidth]{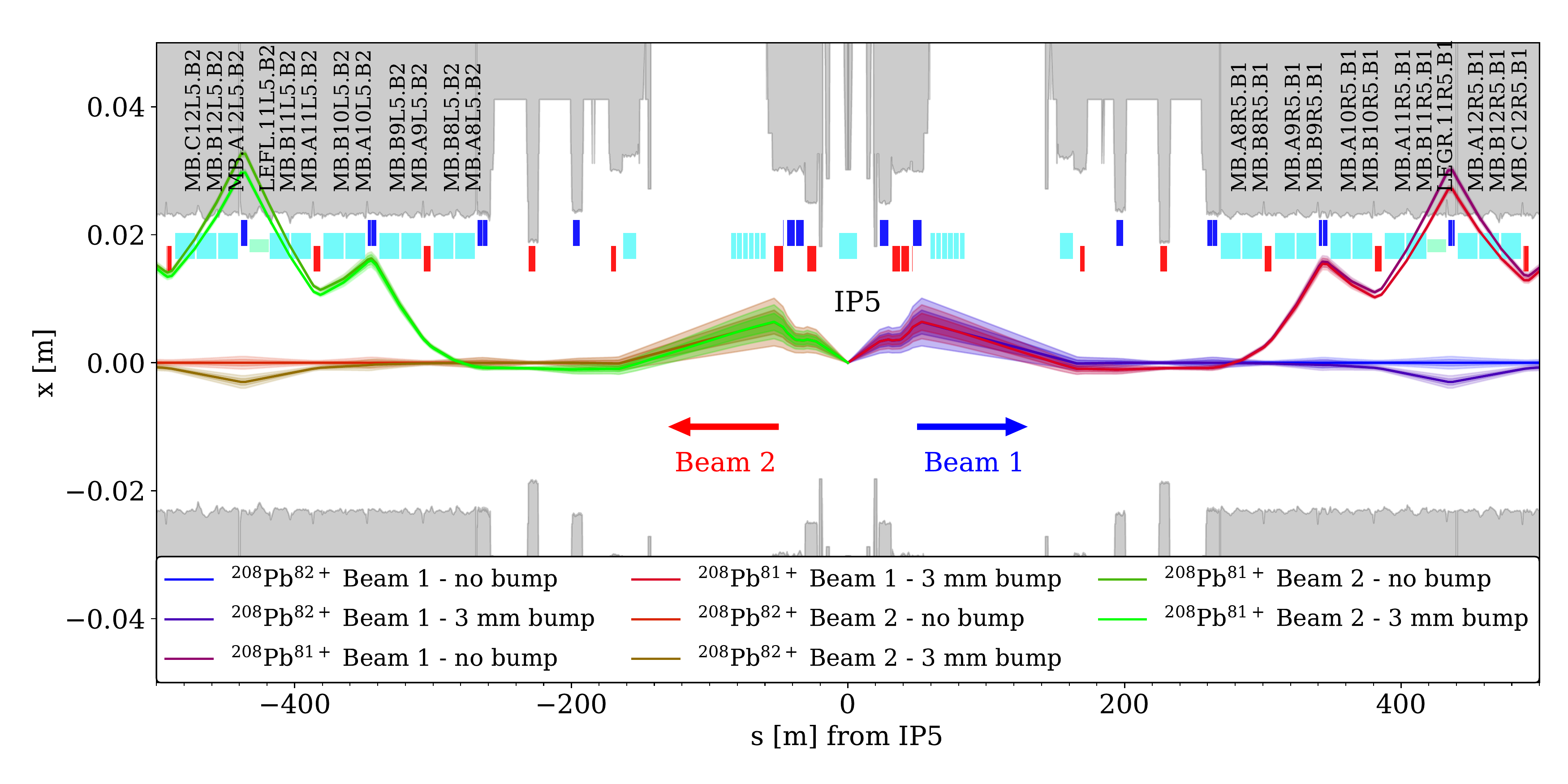}
\vspace{-0.8cm}
\caption{Full view of IR5 and adjusted DS. 
The incoming beam envelopes,  before the collision at IP5, are not shown.  
The grey shaded area represents the aperture, the coloured rectangles on the top the beam-line elements. The effect of a \qty{-3}{mm} orbit bump around Q11 (at $s=\qty{\pm440}{m}$) on the main (\Pb) and BFPP (\PbBFPP) 1$\sigma$ beam envelopes is shown.  Note that the origin of all beams lies at IP5 (center of the plot) such that Beam~2 travels to the left and Beam~1 to the right.
}
\label{f:fullViewIR5}
\end{figure*}
%%%%%%%%%%%%%%%%%%%%%%%%%%%%%%%
\begin{figure}
\centering\includegraphics[width= \columnwidth]{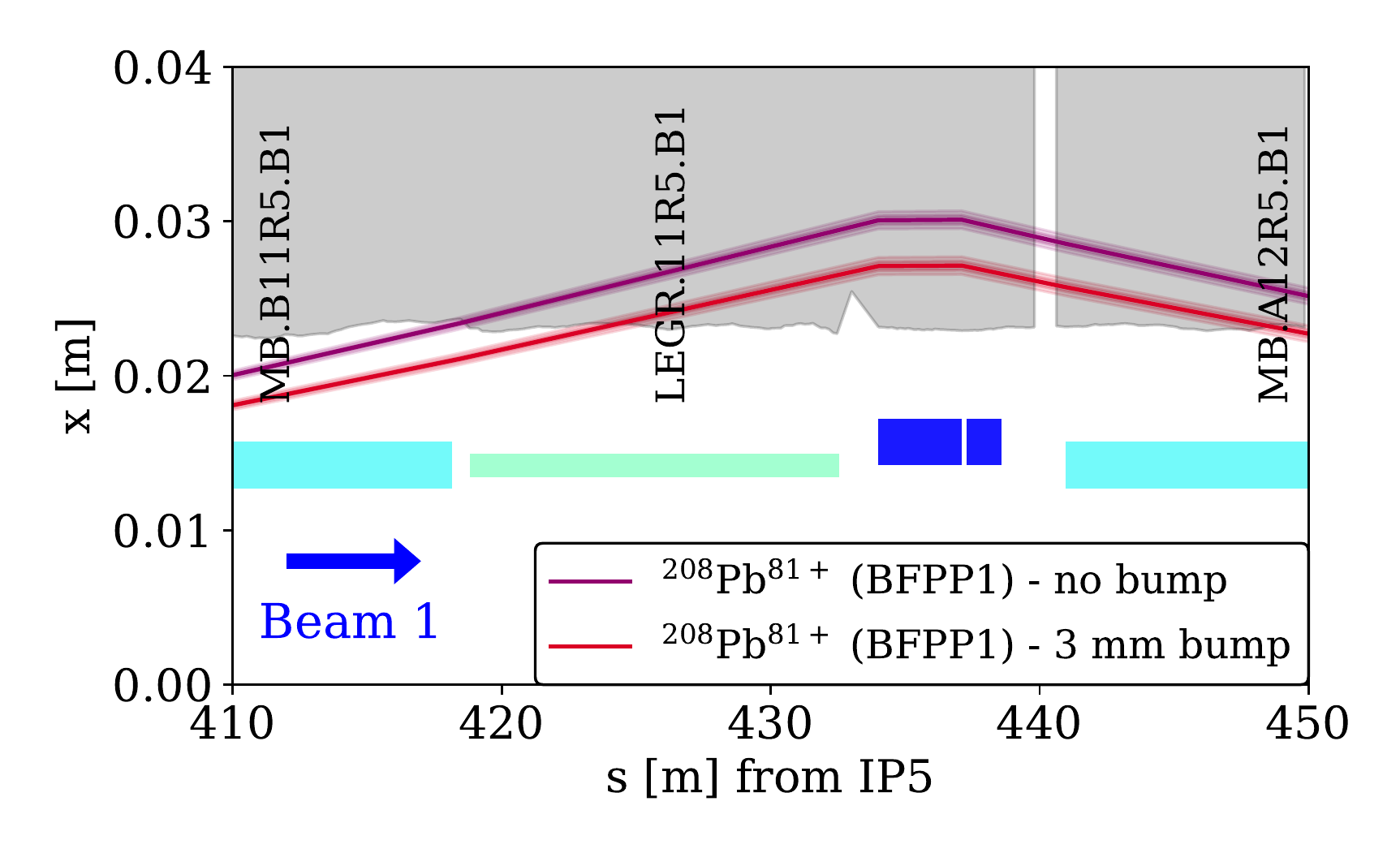}
\vspace{-0.8cm}
\caption[Zoom to impact location of the BFPP beam right of IP5. ]{Zoom into Fig.~\ref{f:fullViewIR5} at the impact location of the BFPP beam right of IP5.  Purple trajectory calculated without orbit bump, red with a bump amplitude of \qty{-3}{mm} at Q11 (blue rectangle).}
\label{f:bfppPathzoom}
\end{figure}
%%%%%%%%%%%%%%%%%%%%%%%%%%%%%%%

%%%%%%%%%%%%%%%%%%%%%%%%%%%%%%%%%%%%%%%%%%%%%%%%%%%%%%%%%%%%%%%%%%%%%%%%%%%%%%%%%%%%%%%%%%%%%%%%%%
%%%%%%%%%%%%%%%%%%%%%%%%%%%%%%%%%%%%%%%%%%%%%%%%%%%%%%%%%%%%%%%%%%%%%%%%%%%%%%%%%%%%%%%%%%%%%%%%%%

\section{Operational Experience in Run~2}
\label{sec:Run2}

\subsection{General Machine Configuration}
In this paper, we focus on the two \PbPb\ operation periods in Run~2.  
In each of them, the IPs of ATLAS, ALICE and CMS were operated with identical $\beta$-functions:  
$\beta^*=\qty{0.8}{m}$ in 2015 and 
$\beta^*=\qty{0.5}{m}$ in 2018 (the design value for ALICE~\cite{lhcdesignV1}), 
and were provided with a similar number of colliding bunch pairs.  
Because of the limit of detector saturation~\cite{alice04}, ALICE was levelled at the design \PbPb\ luminosity
\mbox{$L=\elumi{\mbox{1}}{27} $}~\cite{lhcdesignV1}. 
 
ATLAS and CMS do not have such a limit and could accept the maximum available peak luminosity. 
Therefore, and thanks to the injector performance well beyond design, 
\mbox{$L=\elumi{\mbox{3--3.5}}{27} $  } 
was achieved in 2015~\cite{ipac2016:PbPb2015}. 
This record was broken again in 2018~\cite{Jowett:IPAC2019-WEYYPLM2}, when the reduced $\beta$-functions and a further improvement in the injector performance~\cite{ipac2017:Bartosik:2289479, ipac2017:Alemany-Fernandez:2289674, Huschauer:IPAC2017-THPAB049, Bartmann:IPAC2017-TUPVA007}, 
including shorter bunch spacing and higher single bunch intensities, 
led to a peak luminosity of 
\mbox{$L=\elumi{\mbox{6.1}}{27} $  } 
in these experiments.
Thus the BFPP1 beams emerging from the left and right of the ATLAS and CMS experiments were carrying a power\footnote{At design luminosity of $L=\elumi{\mbox{1}}{27} $, \mbox{$\PBFPP = \qty{26}{W}$.}} of up to \mbox{$\PBFPP \approx \qty{140}{W}$}, which is, as we shall  show later, enough to provoke a quench.

For LHCb, 2015 was the first year of \PbPb\ data taking
and they were provided with only a few tens of colliding bunch pairs. 
Around LHCb, the optics was similar to that of the preceding \pp\ run, with $\beta^*=\qty{3}{m}$. 
In 2018, the $\beta$-function in LHCb was reduced to $\beta^*=\qty{1.5}{m}$. 
In the second half of the 2018 run, 
a new beam production scheme  allowed the bunch spacing to be reduced from  
$S_b=40\lRF\simeq\qty{100\, c}{ns}$, 
to
$S_b=30\lRF\simeq\qty{75\, c}{ns}$, 
in terms of the RF wavelength $\lRF\simeq\qty{2.5\,c}{ns}$. 
In consequence,  LHCb naturally received about 10 times more 
collisions\footnote{
    The IP of LHCb is displaced by 
    $15\lRF\simeq\qty{37.5\,c}{ns}$ 
    with respect to the symmetry point at $s=7C/8$, where 
    the LHC circumference $C=35640 \lRF$. 
    In ``natural'' filling schemes of the LHC, by quadrant, each bunch is located at   
    $s=j C/4+ m S_b$, for   
    $0\le j<4$, $
    0\le m <C/(4 S_b)$, 
    behind the leading bunch. 
    Then encounters occur at 
    $s= k C/8 + n S_b/2$
    for  
    $0\le k<8$, 
    $0\le n <C/(8 S_b)$,
    which always includes the locations of ATLAS, ALICE and CMS at $s=0,C/8,C/2$ 
    but \emph{not necessarily} that of LHCb at $s=1039 C/1188$.
    Bunch spacings of 
    $S_b=10\lRF\simeq\qty{25}{ns}$ (used for \pp) and 
    $S_b=30\lRF\simeq\qty{75}{ns}$ 
    naturally provide   a large number of collisions to LHCb. 
    For the previous Pb beam spacing of
    $S_b=40\lRF\simeq\qty{100}{ns}$, 
    or the future 
    $S_b=20\lRF\simeq\qty{50}{ns}$,  
    some injected bunch trains have to be  displaced from the natural positions  to obtain any collisions at all in LHCb.  
    This generally deprives the other experiments of some collisions. 
}. 
Even though higher  peak luminosities would have been possible, beams were levelled at  
\mbox{$L=\elumi{\mbox{1}}{27} $} (as in ALICE) 
to reduce the risk of quenches (see below).

\subsection{BFPP Orbit Bumps}
\label{sec:OrbitBumpTech}

The importance and consequences of the secondary beams were only realised in their entirety after the final layout of the LHC had been defined (around the year 2000) and no potential counter-measures could be implemented into the cold sections of the accelerator lattice~\cite{lhcdesignV1} before the start-up. 
Early calculations   estimated that BFPP losses would be able to quench magnets already below the nominal luminosity~\cite{epac2004:Jowett:2004rd} and mitigation measures using the deflection of the secondary beams by means of orbit bumps were investigated~\cite{PhysRevSTAB.12.071002} well before the first heavy ions circulated in the LHC.

\subsubsection{The Technique}
The dispersion suppressor (DS) is the lattice section that directly connects the 
regular FODO cells in the arcs to the straight sections on either side of each IP. 
Each DS accommodates four superconducting quadrupoles and eight superconducting dipoles arranged in four cells, numbered 8-11 (see illustration of beam-line elements in Fig.~\ref{f:bfppPath} or \ref{f:fullViewIR5}). 
The last cell (11) is longer than the previous three and contains an extra drift space of approximately the length of a dipole magnet (``missing dipole''). 
This space contains a connection cryostat (labelled LEGR or LEFL to the right or left of the IP)  
that bridges the vacuum, electrical and cryogenic systems to the first optically periodic arc cell. 

Thanks to the providential combination of lattice arrangement and optics  around IP1/5, the impact location of the BFPP beam (see Figs.~\ref{f:fullViewIR5} and \ref{f:bfppPathzoom}) naturally lies in the DS at the end of the second superconducting dipole of the $11^{th}$ cell (corresponds to the $8^{th}$ dipole in the DS, labelled MB.B11) downstream the IP. 
This is the dipole that is followed by the empty connection cryostat without magnet coil. 

This situation allows the use of a horizontal orbit bump around the impact location that pulls the secondary beam away from the aperture just enough to move the beam losses out of the dipole and into the connection cryostat\footnote{
    Note that this orbit bump will also deflect the main beam towards the opposite side of the beam pipe by a maximum of the bump amplitude, usually around \qty{3}{mm}, without compromising machine protection.
    }. 
This is beneficial since the superconducting bus bars in the connection cryostats, which  connect the DS and arc magnets in series, have a much higher steady-state quench level than the magnet coils themselves. 
A rough estimate lies around \qty{200-300}{mW/cm^3}~\cite{QP3, verweijPrivate} at \qty{7Z}{TeV} instead of tens of \qty{}{mW/cm^3} for the magnets. 
Although this  estimate is  very rough, the power deposition is much lower (see Section~\ref{sec:Fluka}) and therefore the risk of quenching is low. 
In addition, the bus bars are located further away from the vacuum chamber and are therefore less exposed to the shower initiated by the impacting ions. 

Figure~\ref{f:fullViewIR5} shows the main and BFPP beam trajectories on both sides of IP5 (Beam~1 travels to the right, Beam~2 to the left). 
Here the natural trajectories are compared to the ones modified by a three-magnet orbit bump with an amplitude of \qty{-3}{mm} around the quadrupole in cell~11 (Q11). 
A zoom in to the BFPP impact location right of IP5 is displayed in Fig.~\ref{f:bfppPathzoom}. 
It is clearly visible that with the orbit bump in place, the impact location is moved from the end of MB.B11 into the connection cryostat.
Since 2015, such orbit bumps have been applied in IP1 and IP5 to mitigate the risk of quenches. 

In IP2 however the situation is different. 
Because  quadrupoles  around this IP have the opposite polarity, the dispersive trajectory of the secondary beams impacts the aperture already at the second dipole of cell~10.
This can be seen in Fig.~\ref{f:bfppPathIP2}, to be further discussed in Section~\ref{sec:IP2strategy}. 
With this optics configuration the BFPP trajectory has a minimum in cell~11 and it is thus not possible to move the BFPP beam into the aperture of the connection cryostat with a simple orbit bump alone. 
In order to safely absorb the BFPP beam here the installation of a new collimator in the connection cryostat in combination with an orbit bump is necessary, as will be further elaborated in Section~\ref{sec:IP2strategy}. 
The collimator upgrade around IP2 is being installed in the current long shutdown.  
Nevertheless the orbit bumps were already applied since 2015 in order to gain experience.

The bump shapes and amplitudes, including the choice of the three orbit correctors building the bump in each IP, were evaluated in MAD-X~\cite{madx} simulations beforehand. 
The exact impact location and angle, and thus the observed loss pattern, strongly depend on the exact beam-line element alignment and aperture.
%, as further discussed in Section~\ref{sec:aperture}. 
Therefore the final amplitude of each bump is optimised empirically during the commissioning phase of each run by a bump amplitude scan. 
This scan aims to move all impacting BFPP particles from the dipole into the connection cryostat. 
For each step in bump amplitude, the loss pattern measured by the Beam Loss Monitors (BLMs)~\cite{holzer05,holzer08a} around the impact locations is compared to the optimal one expected from FLUKA~\cite{fluka_BOHLEN2014211,BATTISTONI201510} simulations (see also Section~\ref{sec:Fluka}).
An example of the behaviour of the BLM signals during such a bump scan and when changing the luminosity is presented in Fig.~\ref{f:blmEvo}. 
These data were collected during the quench limit experiment and will be discussed in detail in Section~\ref{sec:quenchTest}.

After the initial setup and optimization of the bumps with low intensity beam, they were implemented in the operational cycle, before its final validation, 
to automatically be put in place before bringing the beam into collisions. 

\subsubsection{Operational Experience with the Bumps}

The bump amplitudes that were used operationally in 2015/18 are listed in Table~\ref{t:OPbumpAmpl}. 
The bumps were used for the first time in 2015 but, based on the quench limit estimates at that time, 
it was thought that they were not yet strictly necessary for the expected luminosity reach around \lumival{3}{27}. 
Therefore, in that year only, 
the computed bump amplitudes were taken at face value. 
The  optimization procedure described above was  carried out very briefly,  mainly to get experience with the new technique. 
In places where the observed loss pattern with the calculated bump amplitude indicated that the BFPP beams were not well placed in the cryostat, a small variation of  bump amplitude was tried, but the calculated values were retained for  physics operation.   

Since it is important for Section~\ref{sec:quenchTest} it should be noted here that already during the setup of the bumps a left-right asymmetry was observed in IP5. 
While the calculated bump moved the losses well into the cryostat on the right (outgoing Beam~1), high losses were still observed in the dipole on the left (outgoing Beam~2) with a similar bump amplitude. 
The main origin of this effect was later identified to be a misalignment of the real aperture compared to the theoretical one, as will be discussed in detail in Section~\ref{sec:quenchTest}.

\begin{figure*}[tbp]
\centering\includegraphics[width= 0.49\textwidth]{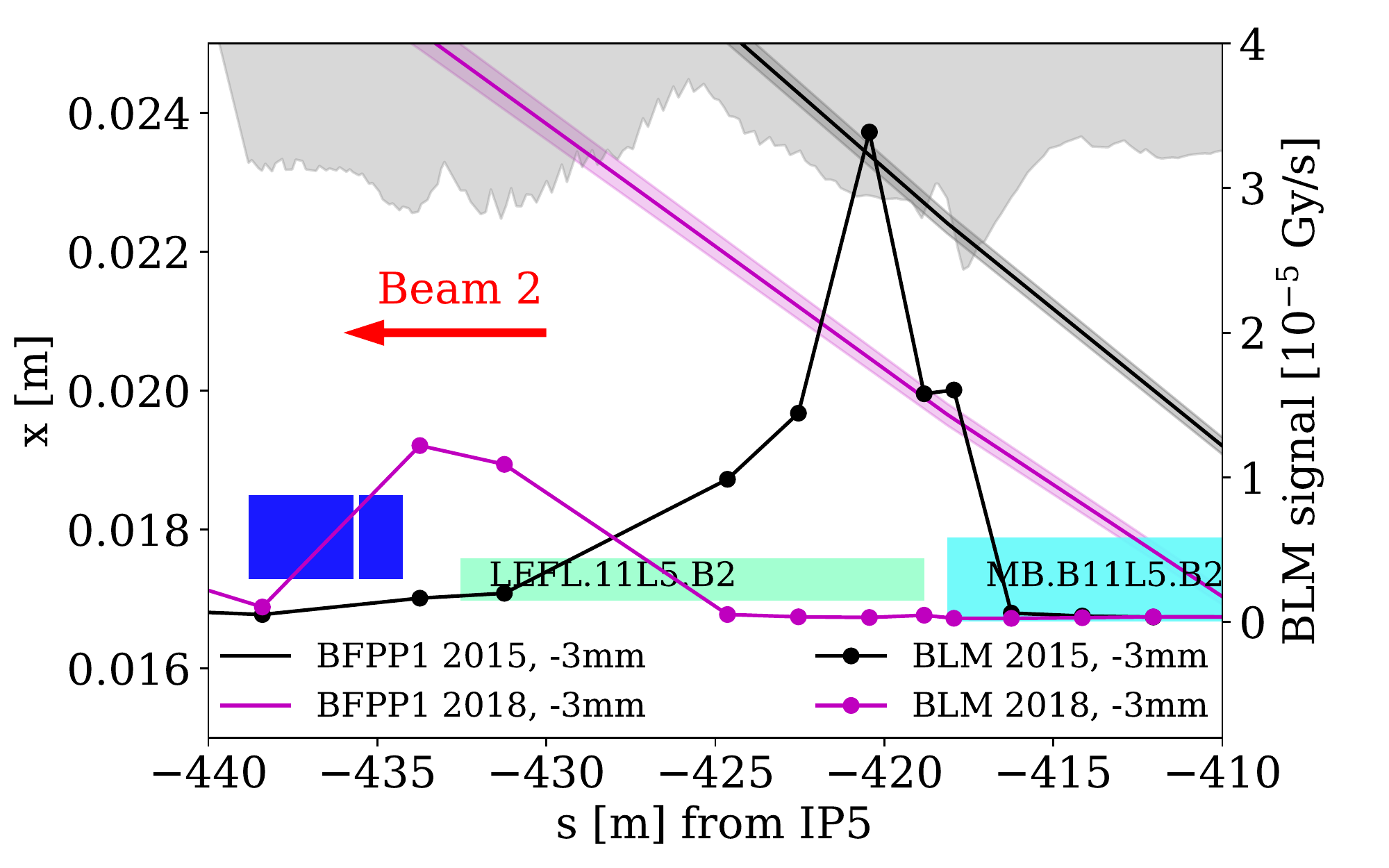}
\centering\includegraphics[width= 0.49\textwidth]{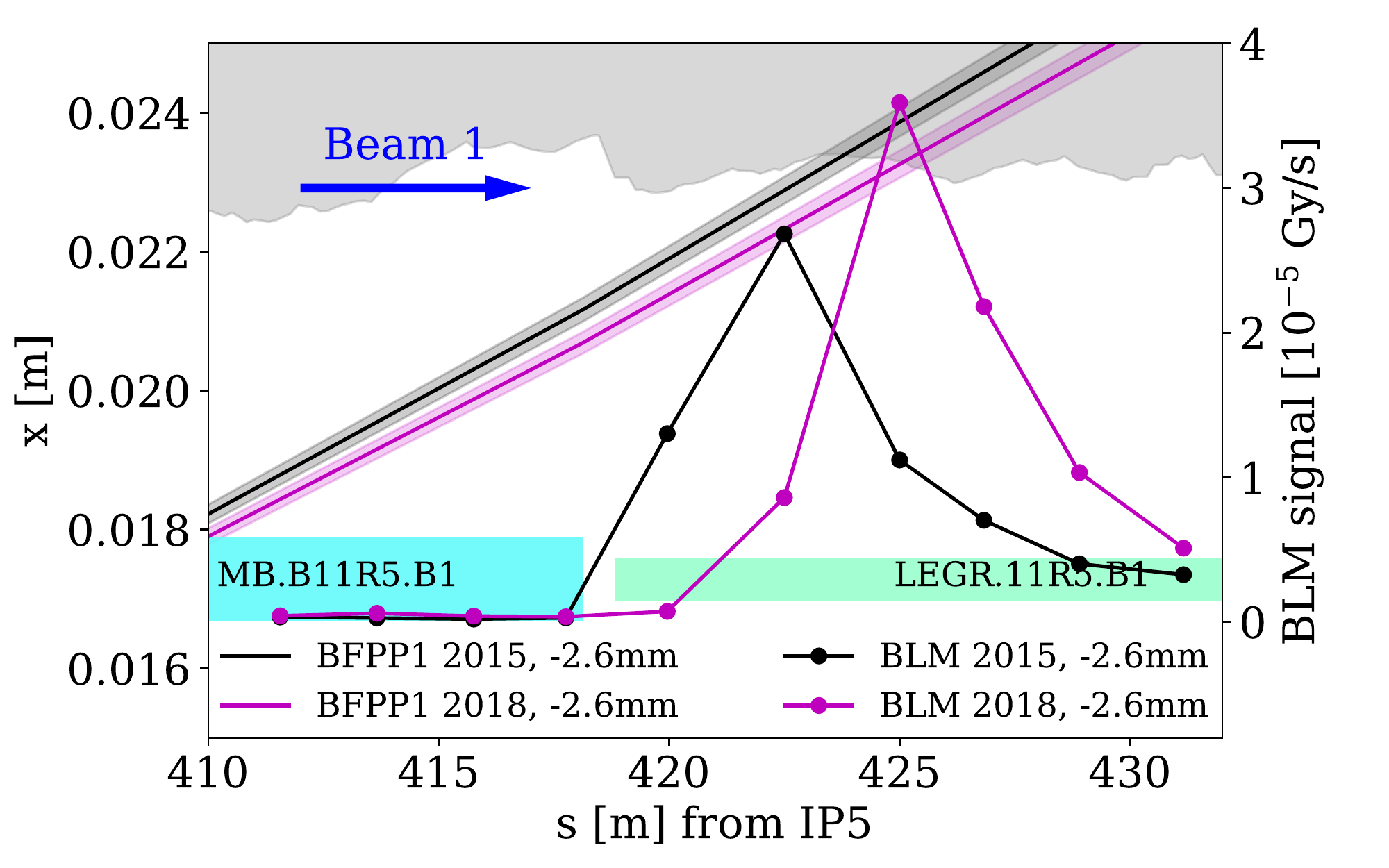}
\caption{ 2015 (black) and 2018 (purple) MAD-X BFPP trajectories (straight lines) and BLM loss signals (corresponding colors) in cell~11 left and right of IP5 for operational bump amplitudes in 2015 (on the left for \qty{-3}{mm} and on the right for \qty{-2.6}{mm} bumps). 
    The rectangles on the bottom correspond to the beam-line elements, the shaded area on the top visualises the reconstructed real aperture. }
\label{f:BLMvsMadxOPBump}
\end{figure*}

With the experience gained from the 2015 luminosity operation and quench experiment, more care was taken to set up the bumps during the commissioning in 2018. 
Detailed bump scans were executed on both sides of IP1, IP2 and IP5 and the empirically optimized  amplitudes were implemented into the cycle. 
Table~\ref{t:OPbumpAmpl} shows that in IR1 the optimal bumps were found to be symmetric, while in IR5 a large left-right asymmetry was still present. 
However, contrary to 2015, a smaller bump was required on the left side of IP5 in order to obtain a loss pattern similar to that on the right side. 
The reason for the smaller bump in 2018 is the sum of two effects. 
\begin{enumerate}
    \item Alignment measurements of the DS elements around cell~11   in 2020 indicated a collective horizontal shift of the beam-line elements towards the outside of the ring of the order of a few hundred micrometers.
    Depending on the real significance of this alignment change, it could lead to an impact position slightly further downstream.
    \item  Nevertheless, the main influence comes from the different IP optics with a smaller $\beta^*$ and larger crossing-angle in 2018. 
    This led to a variation of the local dispersion just behind the IP and a different deflection of the BFPP particles, shifting their impact location   further downstream (see Fig.~\ref{f:BLMvsMadxOPBump}).
     On the left of IP5, the difference from 2015 is of the order of a few meters, while on the right the effect is much reduced. 
     The left-right asymmetry arises because the lattice and optics symmetry between the outgoing Beam~2 on the left and outgoing Beam~1 on the right is not perfect. There are small differences in the matching quadrupole strength and locations between each side of the IP.
\end{enumerate}

%%%%%%%%%%%%%%%%%%%%%%%%%%%%%%%
%%%%%%%%%%%%%%%%%%%%%%%%%%%%%%%

Already in 2015, it was estimated that the   levelled luminosity of \lumival{1}{27} in IP2   
was probably too low to quench a dipole with the BFPP beams.
However, in the absence of experimental confirmation at the time,
the bumps were nevertheless designed and implemented in IR2.
As   will be further detailed in Section~\ref{sec:IP2strategy}, the bumps in IP2 could  only distribute the full load of losses over two cells, rather than move them into the connection cryostat. 
This turned out to be adequate in Run~2, but will  not be enough to mitigate the quench risk for the much higher luminosity that will be provided to ALICE after its upgrade (see Section~\ref{sec:quenchTest}). 

No bumps have been implemented in IR8. 
For the low number of collisions in 2015, losses naturally stayed below the quench limit. 
With the \qty{75}{ns} bunch spacing in the second half of the 2018 run, the 
potential luminosity in IP8 became comparable to the other experiments. 
Since the local geometry and impact distribution are different from IR1/5, 
it cannot be directly assumed that the same power deposition and luminosity limit experimentally found in 2015 for IR5 (see Section~\ref{sec:quenchTest}) applies also to IR8. 
Therefore, LHCb was conservatively levelled at the same value as ALICE (\lumival{1}{27}) 
to protect from quenches as well as share luminosity. 

The bumps have proven to be very efficient and allow at least peak luminosities up to \lumival{6}{27} (measured in 2018) in IP1/5. 
So far no luminosity production fill  has been interrupted by a quench 
or abort (beam dump) due to BFPP losses.

\begin{table}
\centering
\caption[Operationally used BFPP bump amplitudes.]{Operationally used BFPP bump amplitudes in millimeters.}
%\begin{tabular}{|l|l|l|}
\begin{tabular}{lcccc}
\hline
\hline
 	        & \multicolumn{2}{c}{\textbf{2015}}	    & \multicolumn{2}{c}{\textbf{2018}}\\
 \textbf{IP}& \textbf{Left}		& \textbf{Right}    & \textbf{Left} & \textbf{Right} \\
\hline 
IP1 & -3.2 & -2.75 & -2.6 & -2.6 \\
IP2 & -3.0 & -3.0 & -2.6 & -2.0 \\
IP5 & -3.0 & -2.6 & -1.6 & -2.5 \\
IP8 & 0.0  & 0.0  & 0.0  & 0.0 \\
%\hline

\hline
\hline
\end{tabular}
\label{t:OPbumpAmpl}
\end{table}

\subsection{Machine Protection Aspects}

In order to ensure the machine safety when operating with these special orbit bumps a number of measures are applied. 
Because of the different loss mechanisms and beam optics in heavy-ion operation, 
as compared to the preceding \pp\ operation,  
the collimation system has to be re-validated and BLM abort thresholds have to be adjusted. 
The abort thresholds are typically set so that they trigger the extraction of the beams before beam-induced quenches can develop.

The  BLM thresholds have to be adapted before ion bunches are put in the machine for the first time in a run. 
In particular, dedicated abort thresholds, derived with FLUKA shower simulations, are implemented for BLMs at BFPP loss locations next to the four experimental insertions. 
Threshold adjustments are also needed in the betatron cleaning insertion and the neighbouring dispersion suppressor to account for the reduced cleaning efficiency compared to protons. 
The thresholds are then empirically fine-tuned during the initial run period based on the measured BLM response. 

As a standard procedure when commissioning a new beam mode or optics in the LHC the collimation hierarchy is re-validated by artificially exciting a low intensity beam to provoke losses that are observed all around the circumference by the BLM system. 
This provides a so-called \emph{loss map} that is used to verify the collimation cleaning efficiency and that the highest loss rates remain confined to the primary collimator locations.

Since 2015, a special set of loss maps and as well optics measurements have been performed with the \emph{maximum possible amplitude} of the BFPP bumps that might ever be deployed during operation to ensure that they generate no unexpected aperture bottlenecks or optics distortions.
Smaller bump amplitudes are considered to be safe if the largest ones are. 
Therefore, the later optimisation of the BFPP bump amplitude, which requires a sufficient luminosity signal, is allowed to set the operational amplitudes to a smaller value.

%%%%%%%%%%%%%%%%%%%%%%%%%%%%%%%%%%%%%%%%%%%%%%%%%%%%%%%%%%%%%%%%%%%%%%%%%%%%%%%%%%%%%%%%%%%%%%%%%%
%%%%%%%%%%%%%%%%%%%%%%%%%%%%%%%%%%%%%%%%%%%%%%%%%%%%%%%%%%%%%%%%%%%%%%%%%%%%%%%%%%%%%%%%%%%%%%%%%%%%%%%%%%%%%%%%%%%%%%%%%%%%%%%%%%%%%%%%%%%%%%%%%%%%%%%%%%%%%%%%%%%%%%%%%%%%%
\section{BFPP Quench Test}
\label{sec:quenchTest}

In order to probe the luminosity limit for Pb-Pb collisions and to better predict future performances, 
a dedicated test was performed in 2015 that used the BFPP1 beam to provoke a controlled beam-induced quench of a bending dipole. 
The goal of the test was to experimentally determine the dipole quench level for steady-state losses at \qty{6.37\,Z}{TeV}.
Other controlled quench experiments, based on different kinds of loss techniques, had been previously performed at \qty{3.5\,Z}{TeV} and \qty{4\,Z}{TeV} in Run~1 \cite{PhysRevSTAB.18.061002}, but some uncertainty remained concerning the expected quench level at higher energies. For steady-state losses at \qty{7\,Z}{TeV}, i.e. at the LHC design energy, the minimum quench power density for main bending dipoles was estimated to be \qty{22-46}{mW/cm^3} according to electrothermal models and cable stack measurements \cite{PhysRevSTAB.18.061002} (as in Ref.~\cite{PhysRevSTAB.18.061002}, the power density is given as an average density across the cable's cross section). At \qty{6.5}{TeV}, i.e. at the beam energy in Run~2 proton operation, the quench level was estimated to be \qty{\sim 40-55}{mW/cm^3}, while at the Pb operation energy of \qty{6.37\,Z}{TeV}, the quench levels were expected to be a few mW/cm$^3$ higher.
At the outset of the BFPP quench experiment, it was not clear if the peak luminosity which could be achieved in 2015 was sufficient to induce a quench.

The BFPP beams can provide a very clean loss scenario compared with other mechanisms that might induce quenches in proton or heavy-ion operation of the LHC.
The power deposition in the magnet coils can be reconstructed with 
FLUKA particle shower simulations which can then be used to benchmark electrothermal models.
Using the BFPP1 beam to induce a quench has the advantage that the impact point in the magnet can be controlled by modifying the orbit bumps.  
In this way, quenches at the end of the magnet, where the complex coil geometry makes it more difficult to reconstruct the power density, can be avoided. 
Furthermore, the power of the BFPP beam is directly dependent on the luminosity, which can be controlled with the beam separation at the IP.
Preliminary results of  this quench experiment were already presented in Ref.~\cite{Jowett:IPAC2016-TUPMW028}.

\subsection{Setup of the Experiment}

The experiment was performed on 8~December 2015, 
with the highest intensity and lowest transverse emittances available at that date 
to maximise the likelihood of a quench. 
The beams were prepared as for a standard physics fill up to the point of being put in collision. 
The average beam parameters at that time, just before starting the experiment, are listed in Table~\ref{t:beamPar}.

The loss location left of IP5 was chosen as most propitious for the experiment as it exhibited the highest BLM signals in the preceding fills and  
the beam impact point lay further inside the dipole in the absence of the orbit bump. 
On the right of IP5 and on both sides of IP1, the beams would impact closer to the end of the dipole or in the interconnect if no orbit bump was applied. 

Before conducting the experiment, the BLM abort thresholds around the impact location in cell~11 left of IP5 were raised to avoid premature beam dumps. 
Further details of the procedure are given in~\cite{BFPP_MDNote}.

 \begin{table}
        \centering
        \caption{Average beam parameters in the BFPP quench experiment.  Errors indicate the standard deviation. Emittances $\varepsilon_{n(x,y)}$  are normalised values.}
        \label{t:beamPar}
        \begin{tabular}{lc} 
        \hline \hline
Fill number   &4707 \\
Ions per bunch $\Nb$ & $\enum{(1.9\pm 0.3)}{8}$ \\
Bunches colliding in IP5 $k_{c5}$    & $418$ \\
Bunch length $\sigma_z$  & $\qty{9.2\pm 0.3}{cm}$ \\
$\varepsilon_{n(x,y)}$ (Beam 1)   & $\qty{(2.2,1.2)\pm 0.2}{\mu m}$\\
$\varepsilon_{n(x,y)}$ (Beam 2)   & $\qty{(2.0,1.7)\pm 0.2}{\mu m}$\\
\hline \hline
\end{tabular}
\end{table}

\begin{figure*}[tbp]
\begin{minipage}{0.49\textwidth}
\centering\includegraphics[width=\textwidth]{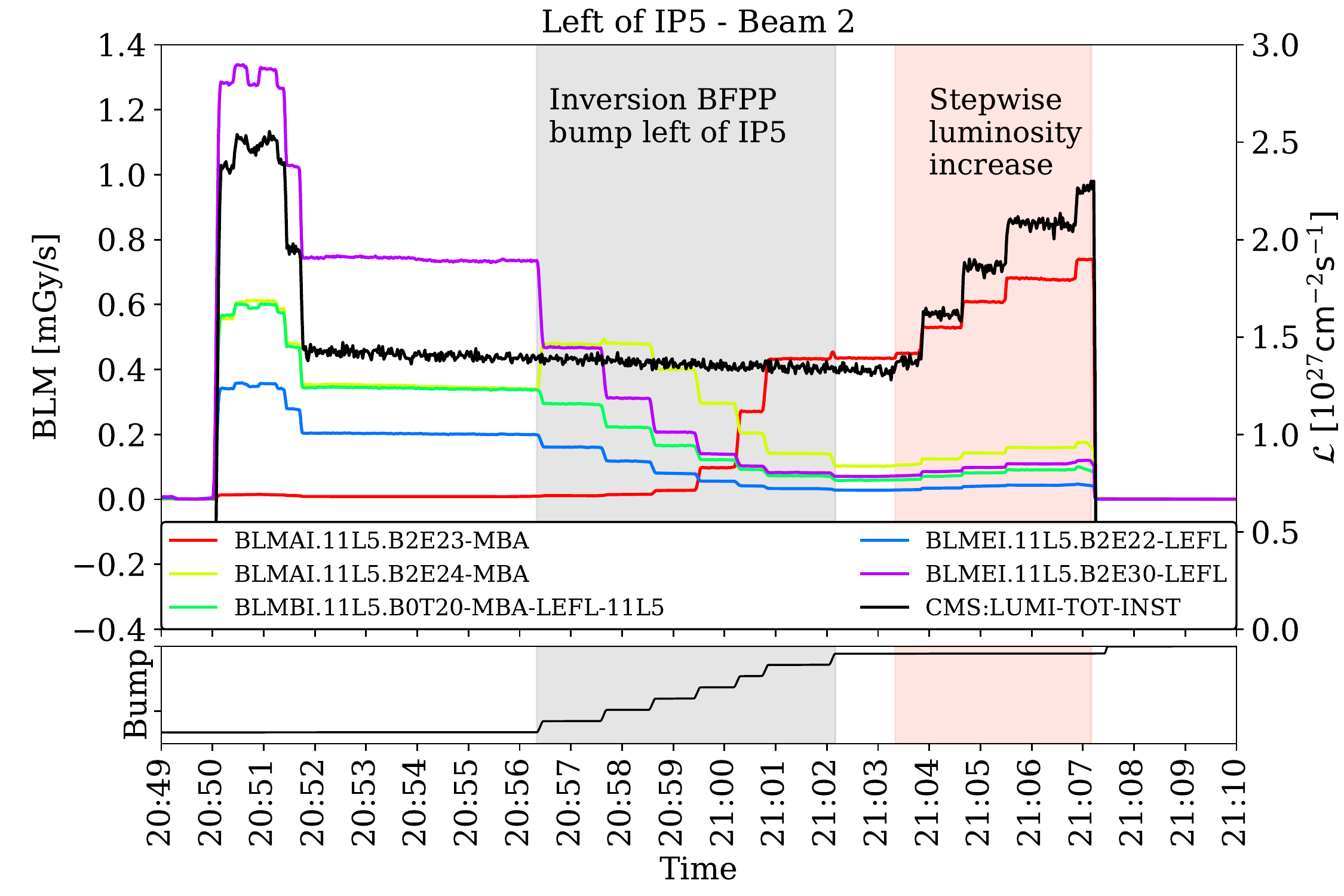}
\end{minipage}
\begin{minipage}{0.49\textwidth}
\centering\includegraphics[width=\textwidth]{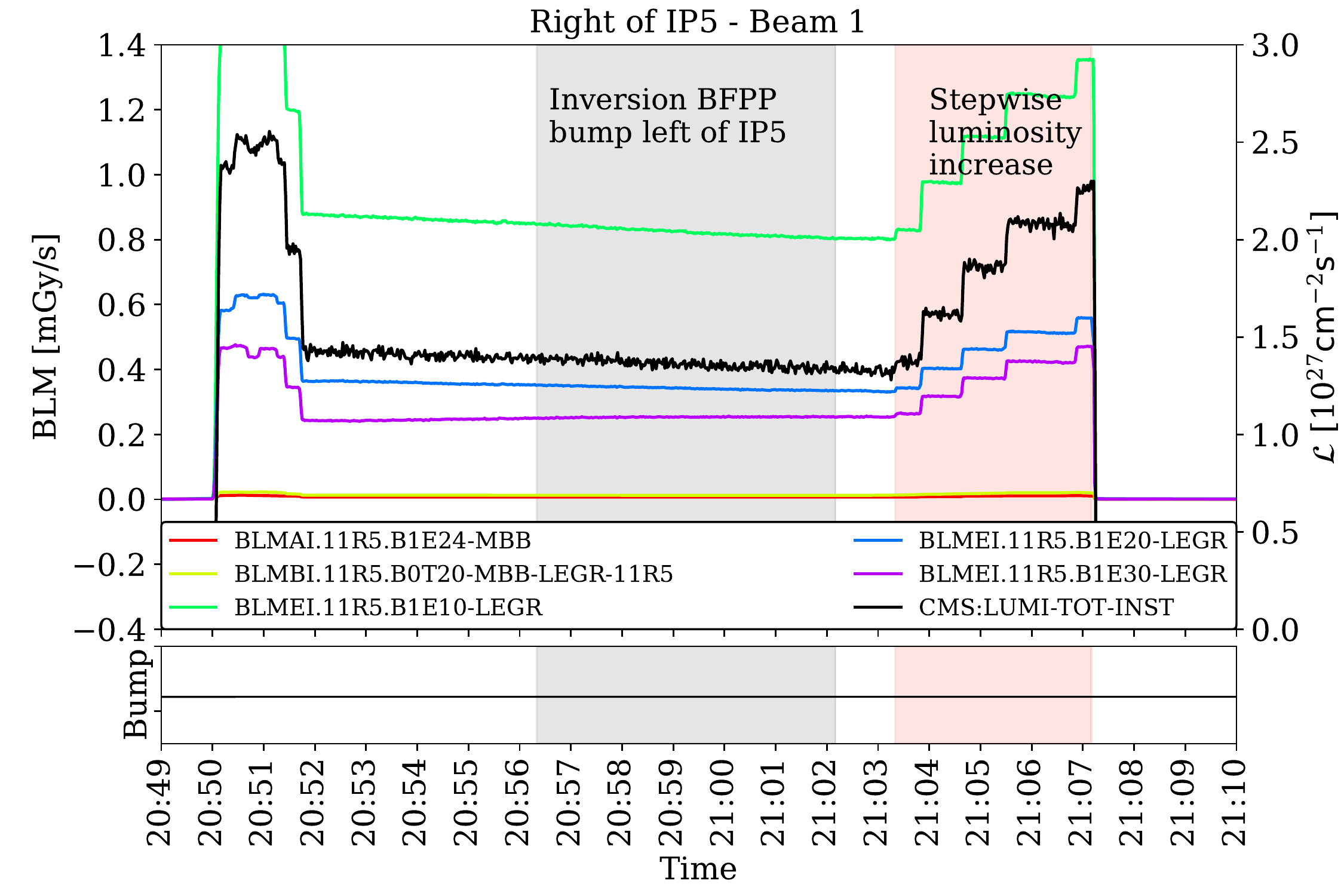}
\end{minipage}
\caption{Evolution of BLM signals around the BFPP impact point  to the left (Beam~2) and right (Beam~1) of  IP5, while stepwise inverting the orbit bump from \qty{-3}{mm} to \qty{+0.5}{mm} on the left of IP5 (grey shadowed period) and increasing luminosity (red shadowed period) during the quench test.
}
\label{f:blmEvo}
\end{figure*}

\subsection{Conducting the Experiment}

Once the beams were colliding, with the BFPP bumps in place as in normal operation, they were re-separated in all IPs in order to reduce burn-off and save peak luminosity for the experiment.
The evolution of the luminosity measured by CMS (black line) can be followed throughout the experiment in Fig.~\ref{f:blmEvo} together with some BLM signals around the BFPP impact location to the left and right of IP5.
The vertical separation at IP5 was reduced sufficiently to discern a pattern on the BLM signals that could later point   clearly to the impact point of the BFPP beam in the bending magnet based on a comparison with FLUKA simulations. 

From here the BFPP orbit bump left of IP5 was reduced from \qty{-3}{mm} through zero and slightly inverted to \qty{+0.5}{mm} (period highlighted in grey in Fig.~\ref{f:blmEvo}), until it was clear that the loss location had moved into the body of the dipole magnet. 
The BLM signals on the left change according to the stepwise reduction of the orbit bump, while on the right they are constant, because the local orbit bump here was not touched. 
On the right the losses with the \qty{-2.6}{mm} orbit bump in place lay inside the connection cryostat (BLMs with names ending on "LEFL" or "LEGR" measure losses inside the connection cryostat; all other BLMs shown measure losses inside the dipole), while on the left losses still occur inside MB.B11. 
Inverting the left orbit bump to \qty{+0.5}{mm} moves those losses  even further upstream into the magnet.

In order to precisely measure the luminosity value leading to the quench, the beam separation at IP5 was reduced in steps of \qty{5}{\mu m}, 
waiting a few minutes at each step for conditions to stabilize (red highlighted period in Fig.~\ref{f:blmEvo}).
After performing the 4th step and arriving at the head on position, a quench of MB.B11L5 developed after around \qty{20}{s} at an instantaneous luminosity of $L\approx \elumi{ 2.3}{27}$ in CMS.

\begin{figure*}[tbp]
\begin{minipage}{0.49\textwidth}
\centering\includegraphics[width=\textwidth]{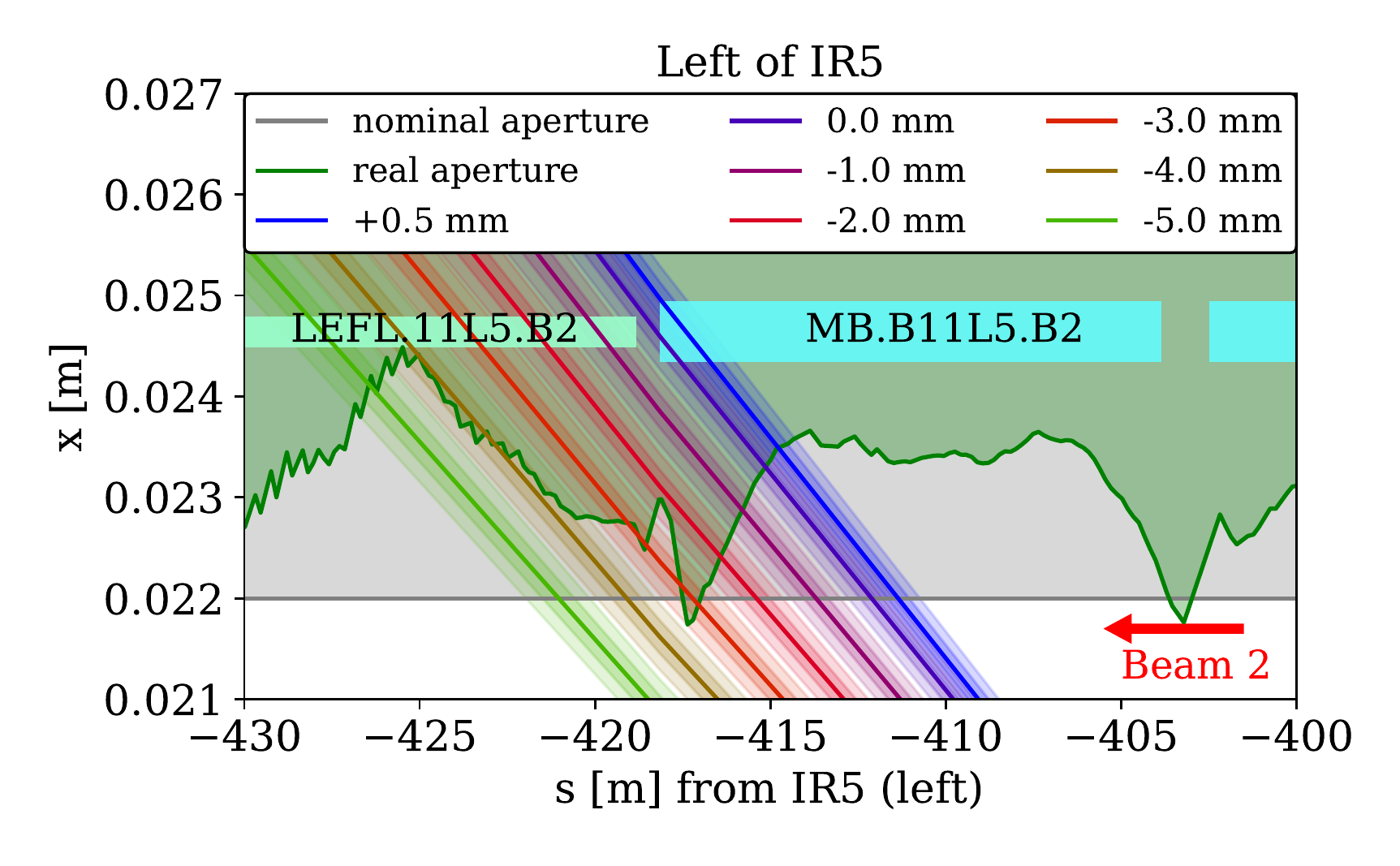}
\end{minipage}
\begin{minipage}{0.49\textwidth}
\centering\includegraphics[width=\textwidth]{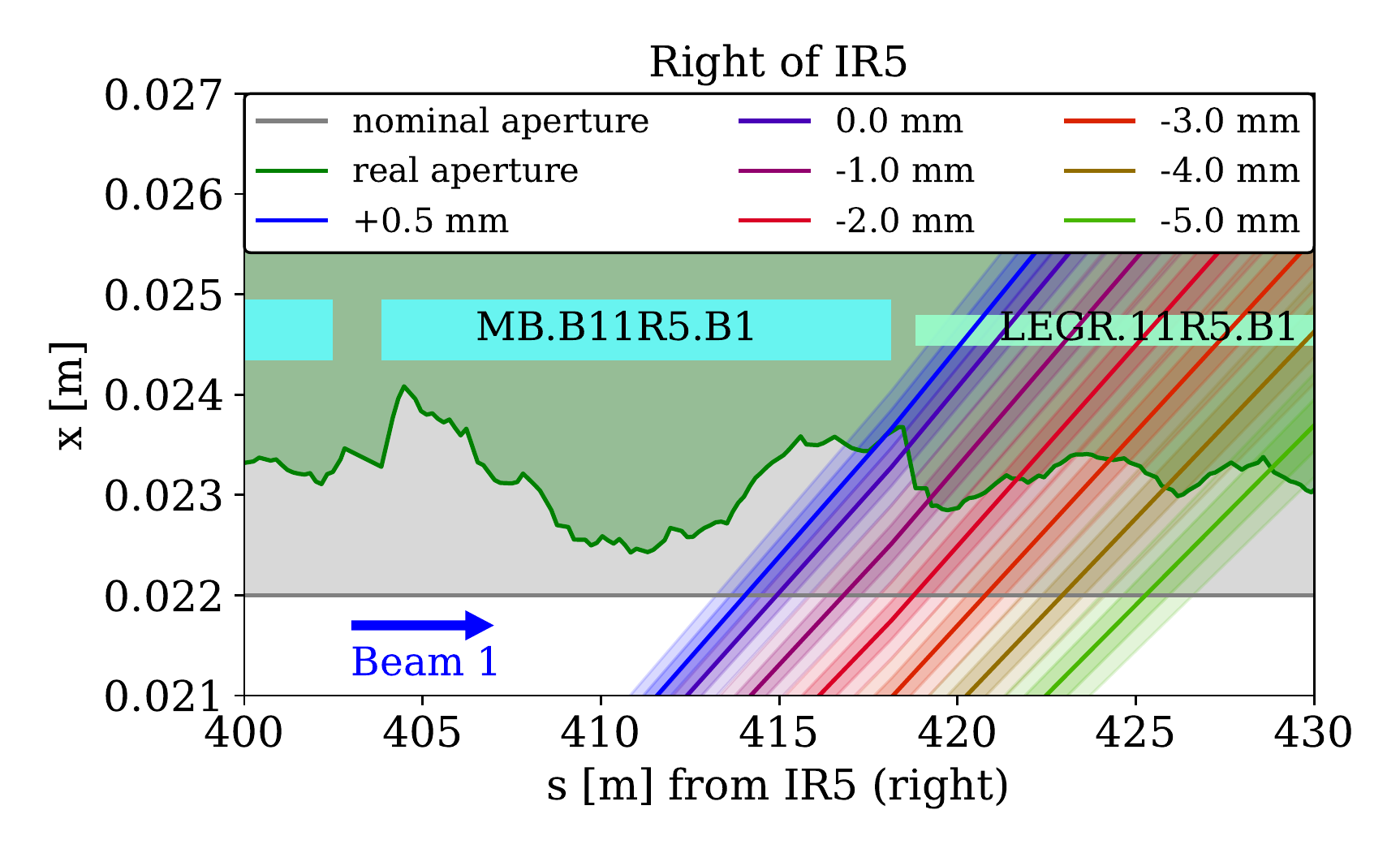}
\end{minipage}
\caption{Standard (\qty{22}{mm}, grey shaded area) and real (green shaded area) horizontal aperture from measurement, superimposed with the BFPP beam trajectory around the impact point left and right of IP5 shown for orbit bump amplitudes between \qty{+0.5}{mm} (blue) and \qty{-5}{mm} (green) calculated for the 2015 operational optics configuration. 
}
\label{f:bfppImpactRealAperture}
\end{figure*}

\begin{figure}[tbp]
\centering\includegraphics[width=0.48\textwidth]{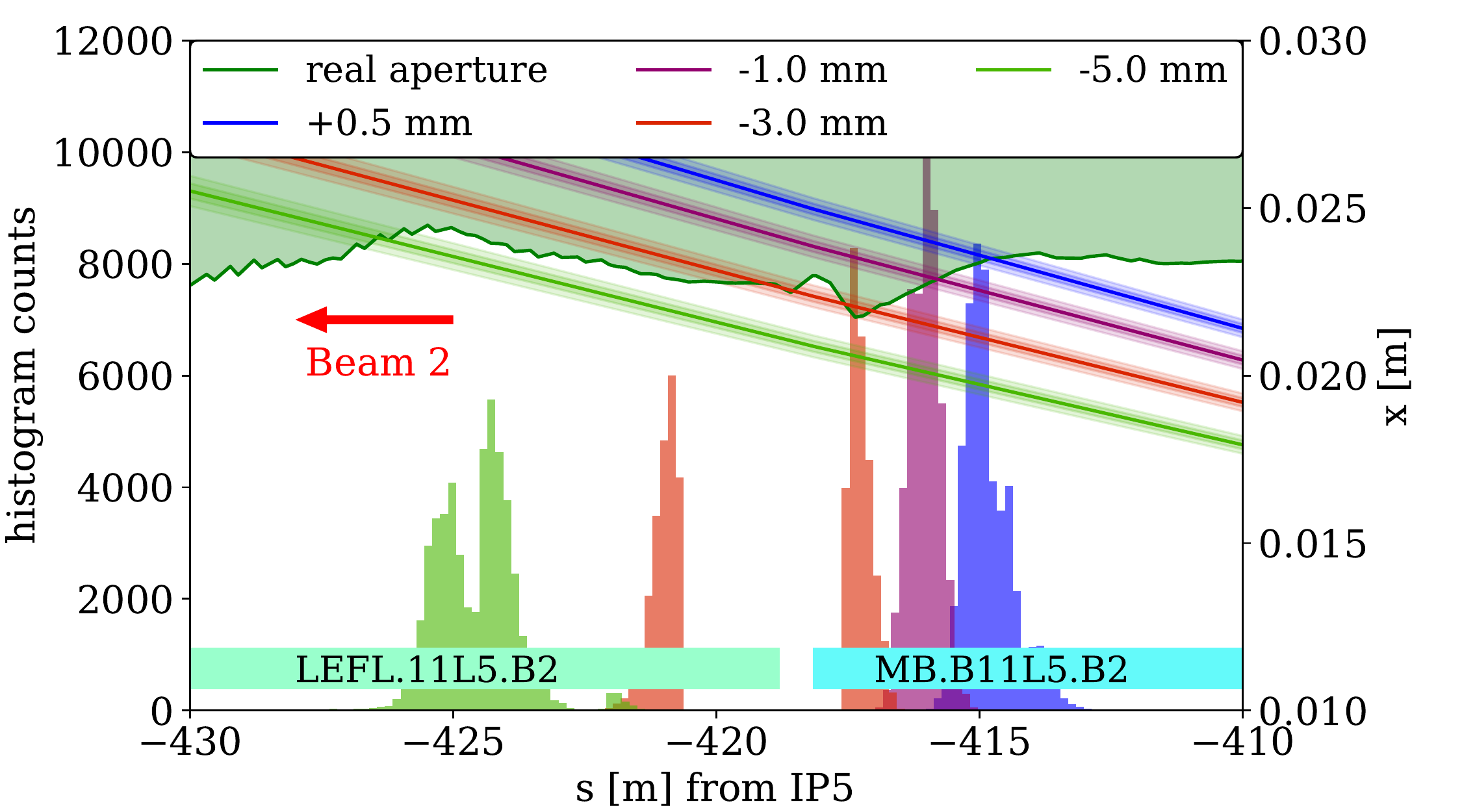}
\caption{BFPP beam distribution  as lost in $s$-direction on the beam screen,  assuming the measured aperture along the impact location to the left of IP5 and the 2015 operational optics configuration with different BFPP bump amplitudes. The transparent rectangles on the bottom indicate the positions of the dipole MB.B11L5 and connection cryostat LEFL.11L5.
}
\label{f:impactDist}
\end{figure}

\subsection{BFPP Impact Location and Distribution}

The exact impact point and distribution depends strongly on the real beam screen aperture and alignment. 
Even a small deviation of the real aperture from the theoretical one will lead to significant differences between observation and simulation of the particle shower, and thus the obtained power deposition and quench limit.
As an example, the ideal mechanical design value of the beam screen aperture is \qty{23.15}{mm}, while the nominal aperture used in the standard LHC \mbox{MAD-X}~\cite{madx} files is reduced to \qty{22.0}{mm}, accounting conservatively for alignment tolerances and mechanical errors. 
Assuming the reduced value rather than the ideal one, shifts the calculated secondary beam impact location several meters upstream.  
Comparing data and simulation of the quench test exhibited non-conformities, which made it necessary to elaborate a more precise description of the real aperture.

During the construction of the LHC, the $x$ and $y$ offsets of the beam screen within each magnet were measured with a longitudinal resolution of \qty{10}{cm} before their installation in the tunnel. 
In order to obtain the real aperture, this offset data has to be superimposed on the ideal beam screen alignment (\qty{23.15}{mm}). 
The difference between the nominal and this corrected  aperture model is indicated in Fig.~\ref{f:bfppImpactRealAperture} for the region around the BFPP impact location.  

Looking at the difference between the grey (nominal MAD-X aperture) and green (real aperture from measurement) shaded areas reveals that the real aperture can feature obstacles, presenting surfaces that are  not parallel to the beam direction to a significant degree. 
While these  may be neglected for other purposes they turned out to be significant for the results of the quench test discussed here.  
On the right, the measured aperture is within the tolerances of the nominal value.  
On the left, however, the tolerance value is exceeded by two spikes, where the one around \mbox{$s=\qty{-417}{m}$} influences the BFPP loss location and distribution for orbit bump amplitudes relevant for the quench test.

The effect of this aperture deformation on the impact distributions in $s$-direction on the beam screen is shown in Fig.~\ref{f:impactDist}. 
As the orbit bump is increased (BFPP trajectories for different bump settings are superimposed), the impact location changes less for a given change in bump amplitude and thus occurs more upstream than expected. 
Further, the shape of the distribution changes significantly  with respect to the nominal case (which can be assumed to be approximately Gaussian). 
This is especially evident around the \qty{-3}{mm} bump (standard value during operation). 
For this setting, the center of the beam hits the tip of the deformation, so that one part of the beam impacts before and another part  after the deformation, leading to two loss peaks with maxima separated by a few meters.  

This also explains the observations made during normal operation.  
For the same bump amplitude, the losses on the left of IP5 were located more upstream compared to the right side, although the optics calculation suggested that the impact locations should have been more symmetric.

%%%%%%%%%%%%%%%%%%%%%%%%%%%%%%%%%%%%%%%%%%%%%%%%%%%%%%%%%%%%%%%%%%%%%%%%%%%%%%%%%%%%%%%%%%%%%%%%%%
%%%%%%%%%%%%%%%%%%%%%%%%%%%%%%%%%%%%%%%%%%%%%%%%%%%%%%%%%%%%%%%%%%%%%%%%%%%%%%%%%%%%%%%%%%%%%%%%%%
%%% This is a copy from the BFPP quench test IPAC paper
%%%% http://accelconf.web.cern.ch/AccelConf/ipac2016/papers/tupmw028.pdf

\subsection{Analysis with FLUKA}
\label{sec:Fluka}

Particle shower simulations were carried out with FLUKA  to  evaluate  the peak power density deposited in the magnet coils during the quench, providing, in turn, a tentative estimate of the steady-state quench level of magnets at \qty{6.37\,Z}{TeV}. 
FLUKA has been benchmarked previously against LHC BLM measurements for different kinds of beam losses \cite{lechner2019-PRAB}; the benchmarks showed that the simulations can reproduce measured signals within a few tens of percent in case of well known loss conditions. 
The studies presented here were based on a similar simulation setup, using a realistic geometry model of the magnet, including beam screen, cold bore, coils, collars and yoke.

To verify the predictive power of the simulation model for the BFPP experiment, simulated BLM signals were compared to measurements. 
The particle shower simulations were based on BFPP1 loss distributions tracked with MAD-X ~\cite{madx}, assuming an orbit bump of \qty{+0.5}{mm}. 
The simulations were first carried out assuming an ideal beam screen model with no manufacturing or alignment imperfections. 
Since such a model cannot reproduce the actual loss location, the impact distribution was artificially shifted in the FLUKA simulations in order to achieve the best match with the measured BLM signal pattern.

\begin{figure}
\centering
\includegraphics[width=0.45\textwidth]{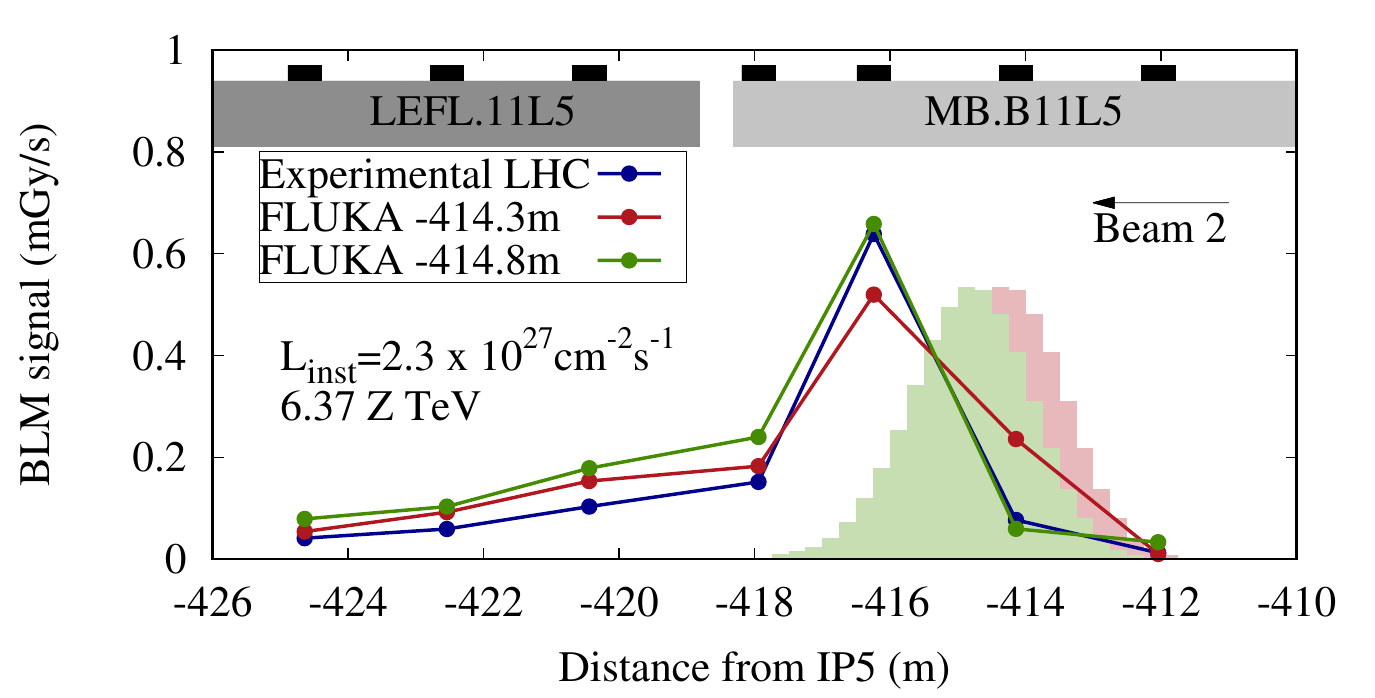}
%\vspace{-10pt}
\caption[BLM signal comparison between experimental data from the quench test and FLUKA data.]{BLM signal comparison between experimental data from the quench test (blue) and FLUKA data assuming two different loss locations (red and green). The particle distributions for the different loss locations assumed in the simulations are shown in histograms of the corresponding color. }
\label{f:flukaResult1}
\end{figure}

Figure~\ref{f:flukaResult1} compares the measured BLM signals and the simulated ones. 
In order to demonstrate the sensitivity of the BLM pattern to the impact location of the BFPP1 beam on the beam screen, the figure shows FLUKA results for two different loss locations differing by \qty{50}{cm}. 
As can be seen in the plot, such a small shift visibly alters the ratio of BLM signals in the vicinity of the loss location. In general, a very good agreement between simulated and measured signals was achieved for an assumed loss location of \qty{414.8}{m} left of IP5. This location is consistent with the loss location predicted by MAD-X if the real aperture model is considered (see Figs. \ref{f:bfppImpactRealAperture} and \ref{f:impactDist}) and therefore provides an independent confirmation about the assumed aperture imperfections.

\begin{figure}
\centering
\includegraphics[width=0.45\textwidth]{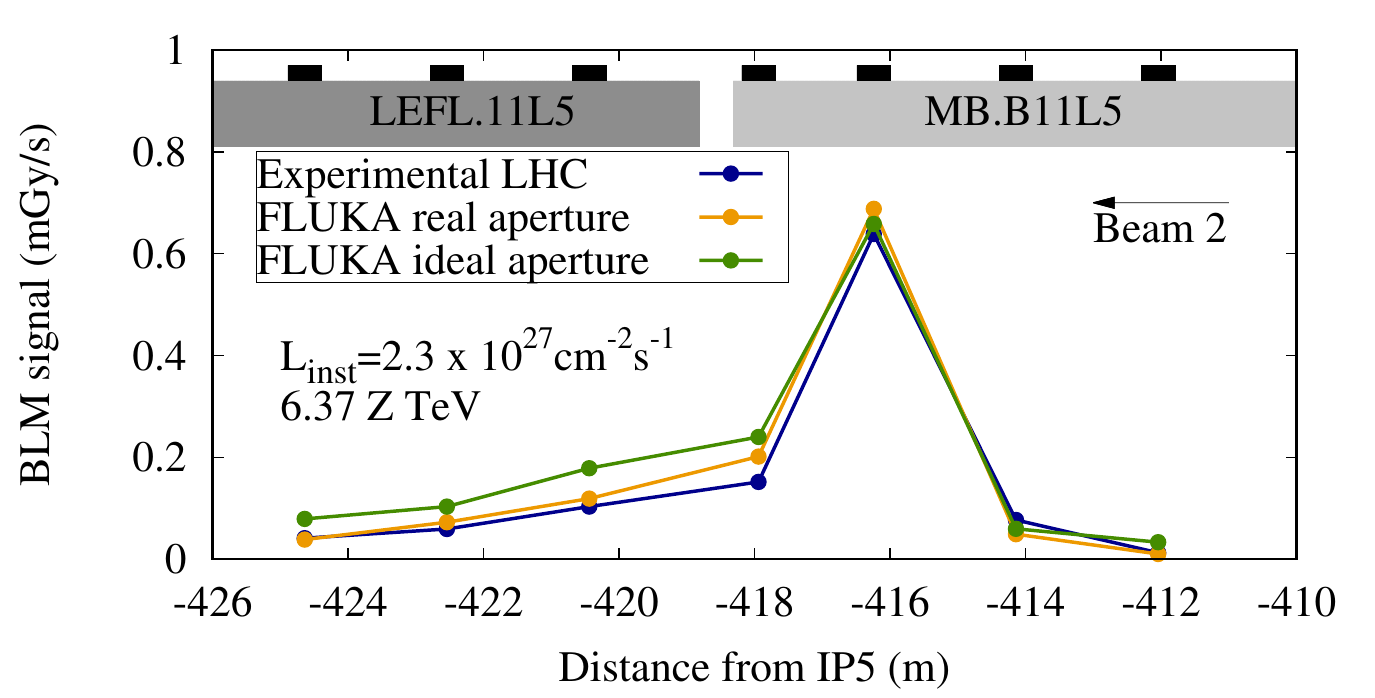}
%\vspace{-10pt}
\caption[BLM signal comparison between experimental data from the quench test and FLUKA data.]{BLM signal comparison between experimental data from the quench test (blue) and FLUKA data assuming both a real (orange) and an ideal (dark green) aperture model. Note that the blue and dark green lines are identical to the ones in Fig.~\ref{f:flukaResult1}. }
\label{f:flukaResult2}
\end{figure}

As can be seen in Fig.~\ref{f:impactDist}, aperture imperfections may not only affect the loss location, but they can also distort the loss distribution of BFPP ions on the beam screen. In order to assess the effect of the distorted impact distribution on the energy deposition in the magnet and BLMs, the ideal beam screen model in FLUKA was substituted by a more realistic one using the real aperture data from Fig.~\ref{f:bfppImpactRealAperture}. This was done by varying the tilt of beam screen model segments (each roughly \qty{1}{m} long) so that they would match the measured aperture data. 
Figure~\ref{f:flukaResult2} compares the simulated BLM patterns obtained with the ideal and more realistic aperture models, assuming in both cases the same location (\qty{414.8}{m} left of IP5).
As can be seen from the plot, the real aperture model produces a closer fit to the experimental BLM signal pattern for BLMs installed further downstream of the peak. The maximum BLM signal is, however, not affected by the aperture model. 

\begin{figure}[tbp]
\centering
\includegraphics[width=0.35\textwidth]{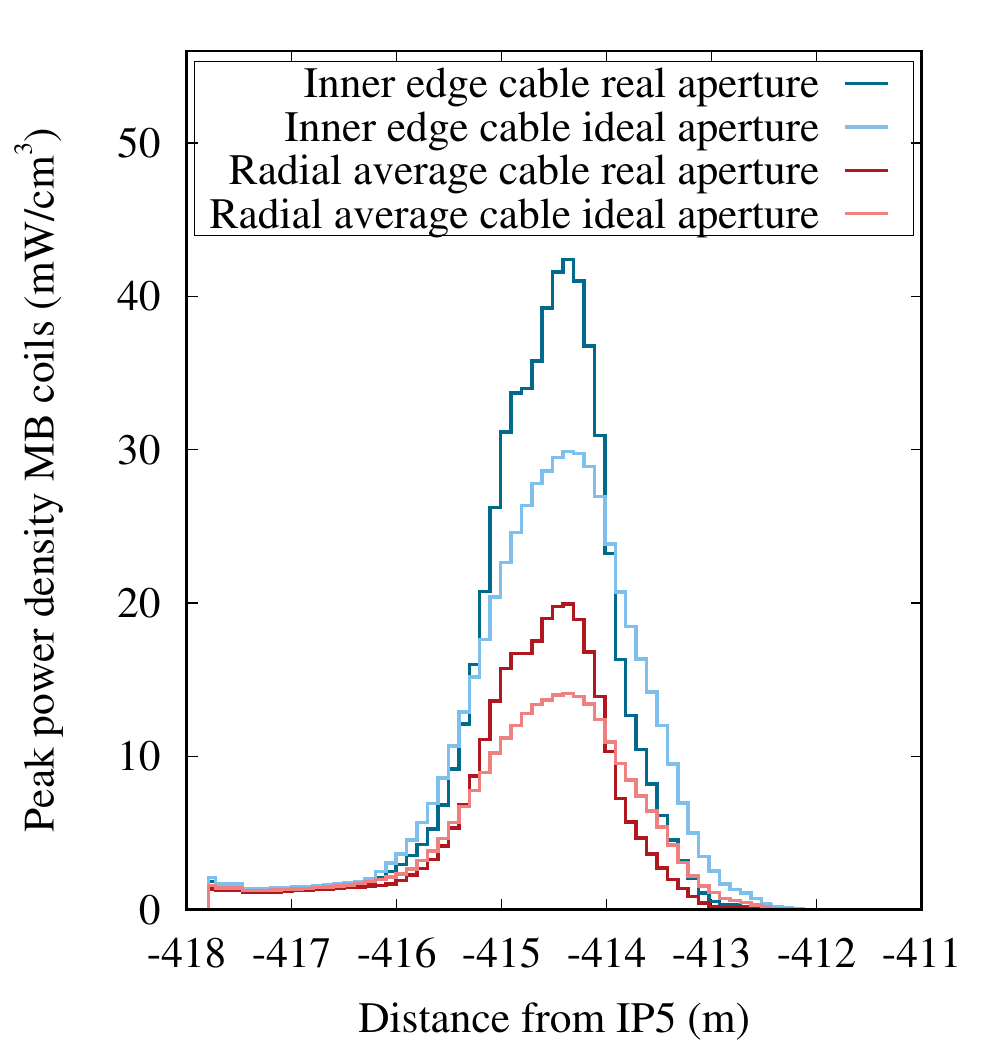}
%\vspace{-10pt}
\caption[Peak power density in the MB.B11L5 coils during the quench test estimated by FLUKA simulations.]{Peak power density in the MB.B11L5 coils during the quench test estimated by FLUKA simulations assuming both an ideal and a real aperture model.}
\label{f:flukaResult3}
\end{figure}

In order to derive an estimate of the peak power deposition in the magnet coils, a cylindrical mesh was placed over the model of the dipole in FLUKA recording the energy deposited in the magnet coils in volume elements  
$ \Delta z \Delta r \Delta \phi =(\qty{10 }{cm}) \times (\qty{0.2}{ cm}) \times (2^\circ)$. 
Figure~\ref{f:flukaResult3} presents the longitudinal distribution of the peak power density in the coils obtained with the real and ideal FLUKA aperture models, respectively.
Both, the peak power density at the inner edge of the cable and the radially averaged density over the cable width are shown. As the heat has enough time to spread across the cables' cross-section, one typically uses the radially averaged power density to quantify the quench level for steady-state losses. The maximum radially averaged power density is estimated to be around \qty{20}{mW/cm^3} in presence of aperture imperfections, while it is around \qty{15}{mW/cm^3} if an ideal beam screen surface is assumed. 
This remarkable difference in the results is much more dramatic than for the BLM signals shown in Fig.~\ref{f:flukaResult2} and can be explained by the proximity of the coils to the beam screen. Because of this proximity, the power density distribution in the coils depends on detailed features of the loss distribution. These features cannot be resolved by the BLMs since they are located outside of the cryostats and are therefore exposed to the far shower tails leaking through the massive magnets. The showers smear out these detailed characteristics of the loss distribution.

Apart from the aperture misalignment, the real loss distribution and hence the maximum power density in the coils depends on the crossing angle, the horizontal and vertical emittance, the momentum spread, imperfections such as small deviations from nominal magnetic field strengths or local inhomogeneities of the beam screen surface at the impact location, but also possible variations of beam and optics parameters. 
Considering these uncertainties, it is estimated that the error on the computed peak power density is at least a few tens of percent, possibly up to a factor of two.

Finally, this could also hint at a non-negligible influence of aperture imperfections in magnet quenches from localized losses. 
If, as our model indicates, BLM signal patterns do not vary widely with aperture imperfections but peak power density deposition does, a greater amount of power density than initially expected could be deposited in the magnet coils due to a local aperture imperfection. This could not be discerned from a real time monitoring of BLM signals during operation, potentially hiding the risk of a magnet quench. 
Beam abort thresholds must therefore incorporate a safety margin.

\subsection{Conclusion from the Experiment}

The reconstructed peak power density in the dipole coils during the quench (\qty{20}{mW/cm^3} when including aperture imperfections) is a factor of two lower than the lower bound of the aforementioned steady-state quench level predicted by electrothermal models (\qty{40}{mW/cm^3} at \qty{6.5}{TeV}). Although the BFPP loss scenario provides 
some of the best conditions to experimentally measure the steady-state quench level of LHC dipoles, it cannot be excluded that a local surface inhomogeneity or another unknown aperture imperfection further distorted the loss distribution in this experiment. 
This could have affected the power density distribution in the coils, but may  not have been visible on the BLMs. A distortion of the impact distribution by an aperture inhomogeneity was also suspected in a previous quench test of an arc quadrupole in Run~1~\cite{PhysRevSTAB.18.061002}. 
In that case, the simulated power density distribution in the coils showed a high sensitivity to a step-like discontinuity in the beam screen surface, which was arbitrarily assumed to be \qty{30}{\mu m} high. 
Although the loss scenario was  different and cannot be compared to BFPP losses, a discontinuity could have likewise increased the peak power density in the present test. 
Uncertainties also remain  concerning the aperture alignment, which had to be reconstructed from several different data sets measured over several years. 
The impact distribution further depends on the instantaneous beam parameters and optics properties. 
It is difficult to exactly quantify these sources of error.

It is also worth noting that during 2015 physics operation it was possible to reach \lumival{\text{3--3.5}}{27} with an inefficient bump, while a quench occurred at \lumival{2.3}{27} during the experiment.
We recall that the loss pattern at the left side of IP5, where the experiment was performed, suggested that the operational orbit bump did not fully move the BFPP beam into the connection cryostat and that the main fraction still impacted in the end part of the dipole.

Because of the sensitivities mentioned, the found peak luminosity leading to the quench cannot directly be assumed to be equivalent for other IPs or under different impact conditions, e.g., new optics or a new impact position within the magnet. 
The results from the quench test were nevertheless used in 2016 and 2018 to adjust the BLM thresholds for heavy-ion operation in all IPs since the test had shown that quenches were possible at lower luminosities than assumed in the original threshold settings. 

To reduce the remaining uncertainty of the steady-state quench level of dipoles, a second quench test in a different location was scheduled in the last few hours of the 2018 run.  
Unfortunately it could not be carried out due to an unexpected failure which 
meant that beams were not available    from the injectors. 
The experiment remains to be repeated in the next \PbPb\ run, currently foreseen at the end of 2022.

\section{Future Heavy-Ion runs}

The constant performance improvement in the injectors and LHC since the first heavy-ion collisions provided the opportunity to briefly operate very close to the HL-LHC target luminosity~\cite{HLLHC_TDR} already in 2018. 
Nevertheless, the 2015 quench test confirmed earlier calculations~\cite{Jowett:619634, PhysRevSTAB.12.071002} that BFPP ion losses would limit luminosity below the HL-LHC target of \lumival{7}{27}, although the limit was found to be higher than originally estimated by these studies. 
Various options have been considered to reduce the power deposition in the superconducting magnets. 
The following section details the mitigation measures planned in all IPs. 
Hardware upgrades are only foreseen in IP2 and  are being installed in the second long shutdown (2019--2021) to be ready for LHC Run~3, when the heavy-ion program enters in its ``high-luminosity'' era.

\subsection{Mitigation Strategy IP1/5: Orbit Bumps}
\label{sec:IP5strategy}

Without the possibility to install special hardware for the alleviation of BFPP beam losses, mitigation measures using the deflection of the secondary particles by means of orbit bumps were investigated well before the first ion run of the LHC~\cite{PhysRevLett.99.144801, PhysRevSTAB.12.071002}. 
The final orbit bump technique,  explained in Section~\ref{sec:OrbitBumpTech}, was  applied already very successfully in LHC Run~2. 
With the possibility to move the impact location of the BFPP ions into the connection cryostat, a peak luminosity over \lumival{6}{27} was reached in IP1/5 without quenching superconducting magnets.

Nevertheless, by moving these losses further downstream, potential risks to other magnets and to the bus-bars in the connection cryostat need to be carefully assessed, in order to avoid new limitations for the future luminosity performance foreseen in the HL-LHC era. 
While the losses in the MB.B11 almost completely disappear with the BFPP bump in place, the adjacent quadrupole MQ.11 (blue rectangle in Fig.~\ref{f:bfppPathzoom}) is exposed to a higher power deposition. 
Simulation studies performed in Ref.~\cite{BahamondeCastro:IPAC2016-TUPMW006} confirm that,
even for HL-LHC specifications of a luminosity of \lumival{7}{27} and at a beam energy of \qty{7Z}{TeV}, 
the peak power density in nearby superconductors, i.e., dipole MB.B11, bus bars of the connection cryostat and the adjacent quadrupole MQ.11, would safely remain below the quench level with more than a factor 10 margin when operating with BFPP bumps. 
Nevertheless, the operational margins for the cryogenic system and radiation effects on electronics remain to be evaluated.

Apart from the power deposition in the magnet coils, the dynamic heat load to be evacuated by the cryogenic system could become a limitation. 
This was also studied in Ref.~\cite{BahamondeCastro:IPAC2016-TUPMW006}. 
According to Eq.~\eqref{eq:PBFPP} the BFPP beams carry about \qty{180}{W} under HL-LHC conditions, 
which is lost in nearby accelerator elements. 
Ref.~\cite{BahamondeCastro:IPAC2016-TUPMW006} estimates that around 75\% of that power goes into the cold mass when the ions are lost deep inside the dipole. 
In the DS region, it is potentially possible to extract \qty{150}{W} (\qty{120}{W} dynamic plus static loads) from cold mass elements at \qty{1.9}{K}. With a higher dynamic load the operational redundancy is jeopardized.
With the orbit bumps in place, more than half of the beam power goes to the connection cryostat, with a large fraction of it (around 45 W) being deposited in lead shielding plates around the vacuum chamber that are thermalized to \qty{50-65}{K}. 
Shifting the losses from the upstream dipole into the connection cryostat therefore also has   a positive effect on the overall load to the cryogenic system.

This, in combination with the operational experience gained at record luminosity close to the HL-LHC target in 2018, confirms that  the orbit bumps a robust solution to guarantee the accessible luminosity for ATLAS and CMS.
Therefore, this technique was confirmed as the HL-LHC baseline strategy for BFPP quench mitigation in these IPs.

\subsection{Mitigation Strategy IP2: DS Collimators}
\label{sec:IP2strategy}

\begin{figure}
\centering\includegraphics[width= \columnwidth]{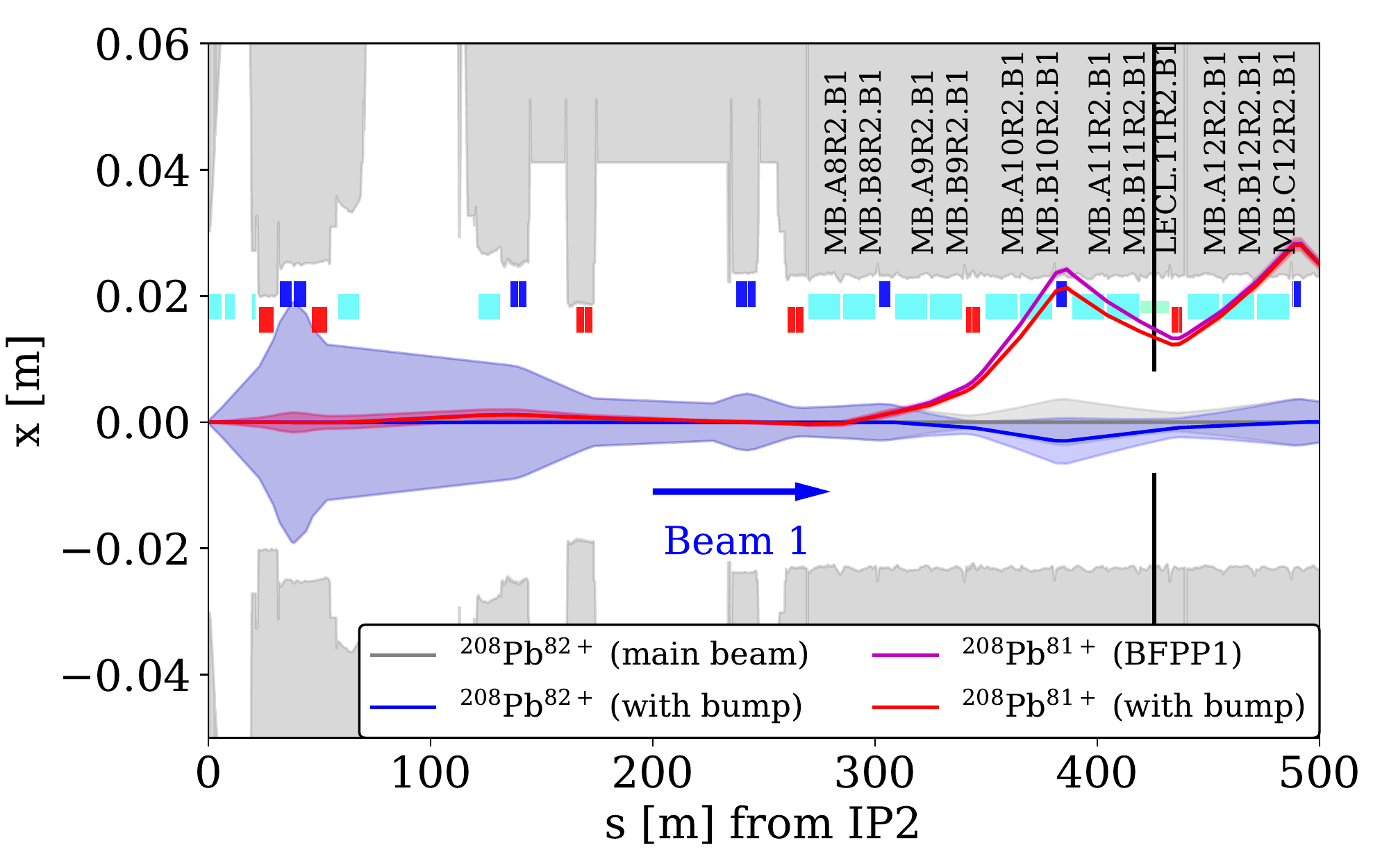}
%\vspace{-0.8cm}
\caption[Main and BFPP1 beam trajectory in Beam~1 direction right of IP2. ]{Main (12$\sigma$) and BFPP1 (1$\sigma$) radial beam envelopes and aperture (grey) in Beam~1 direction right of IP2 (at $s= 0$). Beam-line elements are indicated schematically as rectangles. Dipoles in light blue, quadrupoles in dark blue (focusing) and red (defocusing).  
The  BFPP1  beam impacts in the aperture of the second superconducting dipole magnet of cell~10. An orbit bump around MQML.10R2 of \qty{-3}{mm} can move the impact out of cell~10, such that the BFPP1 beam can be absorbed in a collimator (indicated as black vertical lines at $s = \qty{430}{m}$) placed in the empty connection cryostat in cell~11. }
\label{f:bfppPathIP2}
\end{figure}

Because of the different optics in IR2, that features the opposite quadrupole polarity, the secondary beams cannot be deflected into the connection cryostat by using an orbit bump.
Here the secondary beams are lost already in cell~10 in MB.B10 ($6^{th}$ dipole in DS). 
Since the periodicity of the dispersion function is shifted one cell downstream, the locally generated dispersion is large enough to move the BFPP ions out of the ring acceptance in its first peak (see Fig.~\ref{f:bfppPathIP2}). 

An orbit bump with its maximum around Q10 could only be used to move some (or all) of the losses to cell~12 (MB.C12)\footnote{The dispersion has a minimum (not a maximum as in IP1/5) around the connection cryostat in cell~11 such that the calculated trajectory (without impact) of the BFPP1 beam is closer to the central main beam orbit in cell~11 compared to cell~10 and 12. }. 
This did not impose limits during the 2015 and 2018 runs because of the levelled luminosity in IP2, which kept the  BFPP beam power below the quench limit.

In order to reach the peak luminosity in IP2 foreseen in HL-LHC, similar to values in IP1/5 when operating with orbit bumps, the installation of an additional collimator in the dispersion suppressor, a so-called TCLD\footnote{
    Naming follows the LHC naming convention. 
    As this device is a collimator in the DS that is responsible for absorbing luminosity products, the letter code is the following: TC = Target Collimator, L = Luminosity, D = Dispersion Suppressor. },  
that can absorb the impacting BFPP ions, is necessary.  
However, due to the compact lattice design of the LHC, no space is available to additionally install such a collimator in the lattice without removing existing equipment. Two options have been proposed in the scope of the collimation upgrade programme for HL-LHC~\cite{HLLHC_TDR,bruce15_ipac_HLlayout,mirarchi16_ipac}.

The first option, which was considered in the past but is now obsolete, is to remove and replace an existing standard main dipole magnet (\qty{8}{T}) with a pair of shorter, higher field (\qty{11}{T}) dipoles, which would create space for a tungsten collimator between them. 
In fact this solution was chosen as baseline for installing TCLDs in the DS adjacent to the betatron collimation region in IR7, where the collimators are needed to avoid quenches induced by fragments leaking out of the collimation 
insertion~\cite{HLLHC_TDR,bruce14-PRSTAB-sixtr, bruce14ipac_DS_coll, lechner14ipac-DS-coll,hermes15_ipac}. 
In IR2, this assembly would have to be installed upstream of the MB.B10 on both sides of the IP in order to intercept the BFPP ions before they impact in cell~10.

The second, adopted, solution does not require new magnets. 
If an orbit bump is used to pull the BFPP particles out of the aperture in cell~10, they can continue to travel downstream until they are absorbed by a collimator installed in the connection cryostat in cell~11. 
Ref.~\cite{BahamondeCastro:IPAC2016-TUPMW006} estimates the power deposition in the surrounding superconductors to be uncritical and states that the absorbing properties of the tungsten jaws also reduces the heat load to be evacuated by the cryogenic system.
The installation of such a devices on the outgoing beams on both sides of IP2 is foreseen during the current second long shutdown (LS2) and will enable the HL-LHC performance reach in \PbPb\ collisions for ALICE from Run~3.

\subsection{Mitigation Strategy IP8: Luminosity Levelling}

Since 2018, LHCb  has  requested luminosities comparable to the other experiments bringing it into a regime where BFPP losses become a concern.

Similarly to IR2, the optics in IR8 do not allow  the BFPP ions to be moved  into the connection cyrostat by means of an orbit bump. 
Even an IR2-like bump that distributes losses over two cells seems inefficient. 
No TCLD installation or other upgrade options are presently foreseen. 
This leaves luminosity levelling to a target safely below the quench limit as the only option for BFPP quench mitigation in IR8.

%%%%%%%%%%%%%%%%%%%%%%%%%%%%%%%%%%%%%%%%%%%%%%%%%%%%%%%%%%%%%%%%%%%%%%%%%%%%%%%%%%%%%%%%%%%%%%%%%%
%%%%%%%%%%%%%%%%%%%%%%%%%%%%%%%%%%%%%%%%%%%%%%%%%%%%%%%%%%%%%%%%%%%%%%%%%%%%%%%%%%%%%%%%%%%%%%%%%%

\subsection{Operation with Lighter Nuclei}

LHC operation with lighter nuclear species has been discussed for many years but has so far not been included in the official planning. 
The great success, and high scientific output, of the very short xenon-xenon  run in 2017~\cite{ipac2018:XeXe} increased the interest in collisions with lighter ions in the experimental community~\cite{WG5-report-2018, HI_BSM_2020}. 

From the point of view of collider performance, especially with respect to secondary beams, the operation with lighter ions would be beneficial. 
Event cross-sections for BFPP (\sigmaBFPP) and other ultra-peripheral interactions strongly depend on high powers of the particle's charge number ($Z$)~\cite{Meier:2000ga}:
\begin{eqnarray}
\sigmaBFPP \propto Z^7
\end{eqnarray}

Since the change in magnetic rigidity in the ultraperipheral processes are significantly different, the considerations relating to impact locations, orbit bumps and collimators discussed above for Pb ions do not carry over directly.  
However, as illustrated by Eq.~\eqref{eq:PBFPP}, the power carried by the secondary beams  in collisions of lighter ions   drastically decreases with respect to  \PbPb\ collisions,   naturally reducing the power deposition in the superconductors of the DS.  
Furthermore, the smaller cross-sections lead to a reduced burn-off rate from such interactions, which results in a longer luminosity lifetime, leaving more ions for hadronic interactions.  
Approximate evaluations of these effects are given in \cite{WG5-report-2018}.

%%%%%%%%%%%%%%%%%%%%%%%%%%%%%%%%%%%%%%%%%%%%%%%%%%%%%%%%%%%%%%%%%%%%%%%%%%%%%%%%%%%%%%%%%%%%%%%%%%
%%%%%%%%%%%%%%%%%%%%%%%%%%%%%%%%%%%%%%%%%%%%%%%%%%%%%%%%%%%%%%%%%%%%%%%%%%%%%%%%%%%%%%%%%%%%%%%%%%

\section{Conclusions}

In heavy-ion operation of the LHC, secondary beams created by bound-free pair production processes in the collision dissipate a significant power in the dispersion suppressor regions around the IPs. 
In the absence of mitigation measures, they would quench various superconducting magnets 
and  limit present and future energies and luminosities to values below those already demonstrated.

The BFPP secondary beams were used to measure the steady-state quench level of the LHC dipole magnets at \qty{6.37\,Z}{TeV} in an experiment performed in December 2015.  
A quench was observed at a luminosity  of $L\approx \elumi{ 2.3}{27}$.  
The corresponding peak power density in the magnet coils was estimated to about \qty{20}{mW/cm^3}.

Since the 2015 \PbPb\ run,    orbit bumps at BFPP loss locations have been routinely used and successfully eliminate the quench risk from the BFPP secondary beams in IP1/5. 
In the 2018 run, record peak luminosities of more than \lumival{6}{27} were reached and no physics fill was interrupted by a quench or pre-emptive abort. 
This demonstrated the robustness of the orbit bump technique and its feasibility for the use in IP1 and IP5 under HL-LHC specifications. 
The installation of new collimators around IP2 will  allow the 
future HL-LHC \PbPb\ target luminosity to be provided to the upgraded ALICE experiment. 
FLUKA simulations, benchmarked among others with the presented quench test, demonstrate that the proposed alleviation techniques are efficient and provide a safety factor of at least 10 beyond  the HL-LHC \PbPb\ luminosity reach. 
These studies underline that, in all three IPs, the dissipation of losses into the connection cryostat provides an even distribution of heat load among the various components, facilitating its evacuation by the cryogenic system. 
Nevertheless, the evaluation of the safety margin with respect to other limitations like cryogenic load and radiation effects on electronics need some further study. 
 
The luminosity reach for the LHCb experiment in IP8 strongly depends on the bunch spacing and sharing of total available luminosity with the other experiments. As no accelerator upgrades are foreseen in IR8 and orbit bumps are ineffective, luminosity levelling to a target safely below the quench limit remains for now the only option for BFPP quench mitigation here.

%%%%%%%%%%%%%%%%%%%%%%%%%%%%%%%%%%%%%%%%%%%%%%%%%%%%%%%%%%%%%%%%%%%%%%%%%%%%%%%%%%%%%%%%%%%%%%%%%%
%%%%%%%%%%%%%%%%%%%%%%%%%%%%%%%%%%%%%%%%%%%%%%%%%%%%%%%%%%%%%%%%%%%%%%%%%%%%%%%%%%%%%%%%%%%%%%%%%%

\begin{acknowledgements}

We thank the LHC operations crews and
many colleagues throughout the CERN accelerator and technology sector for their support.  
We are grateful to D.~Missiaen and M.~Giovannozzi for advising us on the accelerator element alignment measurements and aperture measurements. 
We acknowledge B.~Auchmann, who provided us with the quench level estimates for ions beams at \qty{6.5\,Z}{TeV}.
\end{acknowledgements}

\bibliography{References}

%\begin{thebibliography}{99}
%\input{References.bbl}
%\end{thebibliography}

\end{document}